\documentclass[pdflatex,sn-mathphys-num]{sn-jnl}% Math and Physical Sciences Numbered 
\usepackage{graphicx}%
\usepackage{amsmath,amssymb,amsfonts}%
\usepackage{amsthm}%
\usepackage{mathrsfs}%
\usepackage{float}%
\usepackage[title]{appendix}%
\usepackage[table]{xcolor}%
\usepackage{textcomp}%
\usepackage{manyfoot}%
\usepackage{booktabs}%
\usepackage{threeparttable}%
\usepackage{capt-of}
\usepackage{array}
\usepackage{algorithm}%
\usepackage{algorithmicx}%
\usepackage{algpseudocode}%
\usepackage{listings}%
\usepackage{multirow}%
\usepackage{tabularx}%
\newcolumntype{C}{>{\centering\arraybackslash}X}%
\newcolumntype{M}[1]{>{\centering\arraybackslash}m{#1}}%
\theoremstyle{thmstyleone}%
\theoremstyle{thmstyletwo}%
\theoremstyle{thmstylethree}%
\usepackage[normalem]{ulem}%
\useunder{\uline}{\ul}{}%
\def\burl#1{\url{#1}}

\begin{document}

\title[Article Title]{Visibility nowcasting in South Korea: A Machine Learning Approach to Class Imbalance and Distribution Shift}

\author[1]{\fnm{Bong Gyun} \sur{Shin}}\email{tlsqhdrbs3@gmail.com}
\equalcont{These authors contributed equally to this work.}

\author[2]{\fnm{Chan Sik} \sur{Lee}}\email{pouljone@gmail.com}
\equalcont{These authors contributed equally to this work.}

\author*[3]{\fnm{Hyesun} \sur{Suh}}\email{jako403@daejin.ac.kr}

\affil[1]{\orgdiv{Department of AI Big Data}, \orgname{Daejin University}, \orgaddress{\street{1007 Hoguk-ro, Pocheon-si}, \city{Gyeonggi-do}, \postcode{11159}, \country{Republic of Korea}}}

\affil[2]{\orgdiv{Department of Statistics and Actuarial Science}, \orgname{Soongsil University}, \orgaddress{\street{50 Sadang-ro, Dongjak-gu}, \city{Seoul}, \postcode{06978}, \country{Republic of Korea}}}

\affil[3]{\orgdiv{College of Artificial Intelligence Convergence}, \orgname{Daejin University}, \orgaddress{\street{1007 Hoguk-ro, Pocheon-si}, \city{Gyeonggi-do}, \postcode{11159}, \country{Republic of Korea}}}

\abstract{Atmospheric visibility is a critical variable for transportation safety and air quality management, however, accurate prediction remains challenging due to the complex interactions between meteorological conditions and air pollutants, as well as the rarity of low-visibility events. This study introduces a machine learning framework to nowcast visibility in six major South Korean cities. To handle the imbalance in the 2018-2020 training data, we applied the Synthetic Minority Over-sampling Technique with Nominal and Continuous (SMOTENC) and Conditional Tabular Generative Adversarial Network (CTGAN). An ensemble approach combining machine learning and deep learning models was then used and evaluated on a 2021 test dataset. The results revealed a marked decline in predictive performance in the test set compared to the cross-validation phase. This degradation was attributed to a distributional shift between training and testing periods, which was quantitatively confirmed by measuring the Wasserstein distance of the most influential feature identified by SHAP analysis. In general, this study presents a methodology that aims to simultaneously address the dual challenges of data imbalance and temporal distributional shifts, and emphasizes the necessity of accounting for evolving external environmental factors when implementing nowcasting models on time-series data.}

\keywords{Visibility nowcasting, Machine Learning, Data Augmentation, Ensemble Learning, Distribution Shift}

\maketitle

\section{Introduction}\label{sec:introduction}

Visibility has long been recognized as a critical variable in various fields, including traffic safety, aviation operations, and environmental risk management. In particular, reduced visibility directly affects aircraft takeoffs and landings, as well as vehicle operations \cite{hu2017, qian2019_fog}. Visibility distance is generally categorized into three levels. A range of 0–1 km is regarded as a severe critical threshold for issuing weather warnings. A range of 1-5 km is generally considered the threshold for low-visibility conditions. A distance of 5 km or more is regarded as a stable condition with no significant impact on traffic \cite{ortega2023, raj2024, ortega2019_ml}.

Accurate visibility nowcasting is essential to ensure the safety of air, sea, and road traffic and to reduce the social costs associated with air pollution. In particular, low visibility events caused by fog or fine dust can cause serious social disruption and economic losses, such as chain-reaction collisions and flight cancellations \cite{taszarek2020, zhai2020, ding2024}. Therefore, it is a crucial task to develop an early warning system that can proactively respond to potential risks by accurately nowcasting visibility in advance, comprehensively considering various meteorological and environmental factors. However, in Korea, the focus remains on improving the performance of the visibility measurement system rather than on nowcasting and forecasting visibility \cite{lee2019}.

To establish such an early warning and response system for visibility, ensuring the accuracy of the nowcasting model is essential. Currently, various approaches are used to predict atmospheric visibility, with statistical models based on meteorological data being the most common \cite{yu2021}. However, numerical-based statistical models have limitations in that they cannot fully explain nonlinear relationships. Therefore, this study aims to overcome these limitations by implementing visibility nowcasting based on machine learning and deep learning. Machine learning and deep learning models can learn nonlinear patterns, allowing them to more accurately describe the complex relationships between meteorological and air quality variables than traditional linear statistical models and have generally shown superior predictive performance \cite{yu2021, kim2022, zhou2024}. As traditional statistical models are inefficient at learning rare weather events (minority classes) in the meteorological field, this study aims to apply machine learning and deep learning-based models \cite{chantry2021, fathi2022, aguasca2019}. Beyond simply applying standard machine learning algorithms, this study introduces a framework designed to tackle two fundamental challenges in environmental time-series modeling: extreme class imbalance and temporal distribution shift. To systematically address these methodological hurdles, the proposed framework incorporates the following core technical contributions:

First, to address Extreme Data Sparsity—where the minority events (Class 0 and Class 1) account for only approximately ${8.9}\%$ of the total dataset on average across all regions—we propose a systematic evaluation of diverse data augmentation strategies. Rather than proposing a singular solution, we explore linear interpolation (SMOTE) \cite{chawla2002}, generative adversarial modeling (CTGAN) \cite{xu2019}, and their hierarchical combination to identify model-specific optimal configurations. This comprehensive assessment reveals that the effectiveness of augmentation varies significantly with model architecture, highlighting the necessity of an adaptive approach rather than a universal solution.

Second, we implement a two-stage soft voting ensemble architecture to proactively mitigate Temporal Distribution Shift caused by annual variations in climate and air quality patterns. Since specific-year biases can distort a single model's decision boundary, our framework aggregates prediction probabilities through model stage ensemble (combining ML and DL models within each fold) and fold stage ensemble (aggregating across cross-validation folds). This mechanism aims to cancel out variance in individual model errors and improve robust generalization performance against evolving environmental conditions.

Third, this study employs SHAP-based feature importance to identify the most influential variable for each region, and then measures the Wasserstein distance---a metric well-suited for comparing distributions with differing shapes or supports \cite{martin2017}---on this feature to quantitatively identify distribution shifts between the training (2018–2020) and test (2021) datasets. By focusing on this high-impact feature, we empirically confirm that the distributional distance to the test data substantially exceeds the internal variation within the training period. This finding directly proves that the observed performance degradation is primarily caused by these validated shifts in high-impact regional variables, providing a rigorous diagnostic basis for our model's behavior.

Finally, to validate the practical utility of our framework, we conduct a comparative analysis against the traditional Logistic Regression model. This comparison serves to ascertain the extent of performance improvement achieved by our final model over established statistical benchmarks, highlighting the advantages of our approach in handling complex, real-world meteorological data.

The findings of this study contribute to traffic safety and environmental risk management not only by optimizing predictive performance but also by providing a rigorous diagnostic methodology to identify the underlying causes of performance degradation in the presence of temporal data distribution shifts. This emphasizes the necessity of continuous monitoring and adaptation to evolving external environmental factors when implementing forecasting models on time-series data.

\section{Methodology}\label{sec:methodology}

\begin{figure}[htbp]
    \centering
    \includegraphics[width=0.7\textwidth]{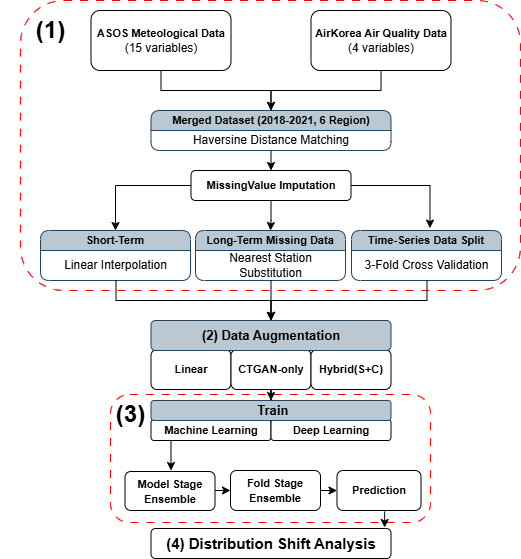}
    \caption{The overall framework of this study. The pipeline consists of four main stages: (1) Data Collection and Preprocessing—merging meteorological data from KMA's ASOS and air quality data from AirKorea, followed by missing value imputation; (2) Data Augmentation—applying SMOTENC and CTGAN to aim to address severe class imbalance in minority classes (Class 0 and Class 1); (3) Model Training and Ensemble—training five models (XGBoost, LightGBM, ResNet-like, FT-Transformer, DeepGBM) via 3-fold cross-validation and combining them through two-stage soft voting ensemble; (4) Distribution Analysis—interpreting the observed performance degradation between validation and test datasets by measuring distributional shifts via Wasserstein distance.}\label{fig:overall_framework}
\end{figure}

This section introduces the general methodology of the visibility nowcasting model proposed in this study. Specifically, it details the data preprocessing steps, including data collection, merging, and missing value imputation, as well as data augmentation methods, and the configuration and ensemble techniques of the machine learning and deep learning models. Figure \ref{fig:overall_framework} shows the overall framework of this study, from data collection to prediction using the final model.

Recently, the impact of various air quality variables, such as fine particulate matter ($PM_{2.5}$) and ozone ($O_3$), on visibility has been increasingly highlighted \cite{deng2008, cheung2011, zhang2020}. Consequently, the importance of $PM_{2.5}$ and $O_3$ has increased in visibility nowcasting models. However, information on $PM_{2.5}$ and $O_3$ is not available in the primary dataset of this study, the KMA's ASOS data, but is included in AirKorea's ground-based air quality data. Therefore, this study used a merged dataset from ASOS and AirKorea data. In this study, we propose the following approach for visibility data from six major cities (Seoul, Busan, Incheon, Daejeon, Daegu, and Gwangju). First, we aim to mitigate class imbalance by artificially increasing smaller Class 0 and Class 1 using the Synthetic Minority Over-sampling Technique with Nominal and Continuous (SMOTENC) and Conditional Tabular GAN (CTGAN). Subsequently, we perform multiclass visibility nowcasting using a total of five models—XGBoost \cite{chen2016}, LightGBM \cite{ke2017}, ResNet-like \cite{he2016}, FT-Transformer \cite{gorishniy2021}, and DeepGBM \cite{ke2019}—and finally apply a soft voting ensemble technique that integrates the prediction probabilities of multiple models to aim to enhance prediction accuracy and stability. In this study, we particularly emphasize the performance of low-visibility detection by using the Critical Success Index ($\mathrm{CSI}$) as a primary metric \cite{joseph1990} to analyze the effectiveness of the augmented data and the synergistic effects of the ensemble methodology. Furthermore, concerning the new issues identified through this series of analyses, we conduct additional analysis to determine their causes by measuring the “distance between distributions” using the Wasserstein distance.

\subsection{Data Preprocessing}\label{sec:data_preprocessing}

\begin{figure}[htbp]
    \centering
    \includegraphics[width=0.5\textwidth]{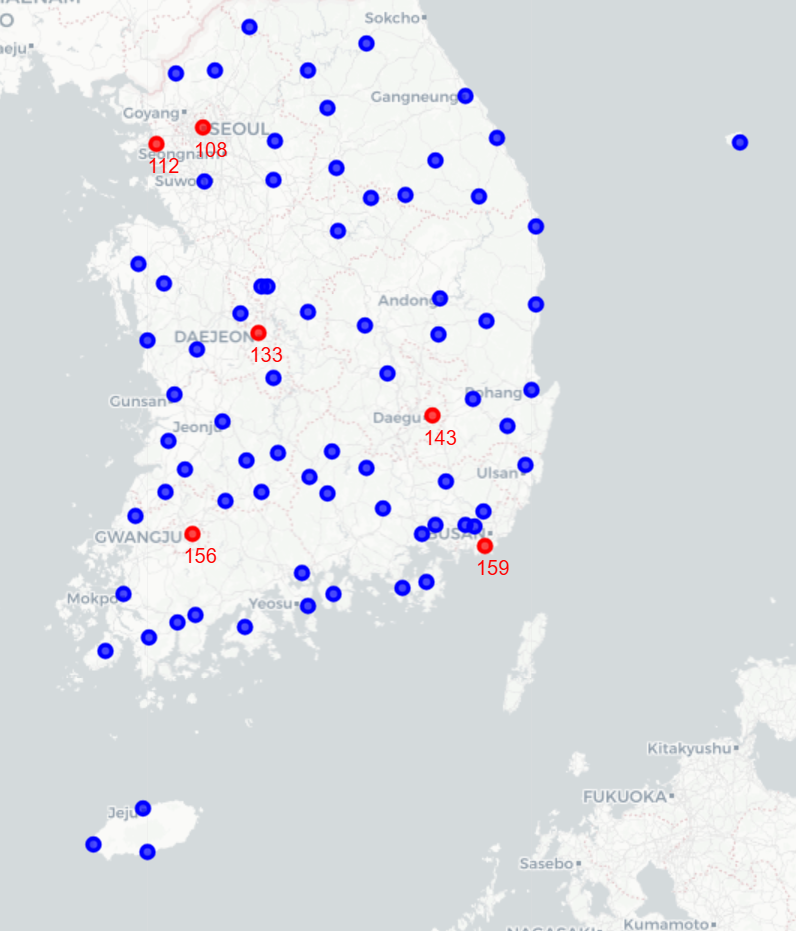}
    \caption{Map of the six major South Korean cities included in the study}\label{fig:study_area_map}
\end{figure}

A total of 98 ASOS observation points and 256 ground-based air quality stations operated by AirKorea are located throughout the country. In this study, ASOS data was utilized as a reference due to incorporation of visibility data, which constitutes the target variable intended for prediction. The objective of this study is to implement a visibility prediction model for the six major Korean cities $(\mathrm{Seoul}_{108}, \mathrm{Busan}_{159}, \mathrm{Incheon}_{112}, \mathrm{Daegu}_{133}, \mathrm{Daejeon}_{143}$ and $\mathrm{Gwangju}_{156})$ by integrating observed meteorological and air quality data from 2018 to 2021, as illustrated in Figure \ref{fig:study_area_map}.

Table \ref{tab:features_summary} lists the variables collected from the KMA's ASOS and AirKorea's ground-based air quality monitoring stations. The concentrations of major pollutants affecting visibility, such as particulate matter ($PM_{10}$), fine particulate matter ($PM_{2.5}$), ozone ($O_{3}$), nitrogen dioxide ($NO_{2}$), carbon monoxide ($CO$), and sulfur dioxide ($SO_{2}$), were collected at 1-hour intervals from 2018 to 2021 \cite{deng2008, cheung2011, zhang2020}.
\begin{table}[htbp]
    \caption{Summary of collected meteorological and air quality variables}
    \label{tab:features_summary}
    \begin{tabularx}{\textwidth}{C}
        \toprule
        \textbf{ASOS Observation Variables (15)}                                                                                                                                                                                                                                                                                                                                               \\
        \midrule
        Temperature ($^{\circ}$C), Wind speed (m/s), Relative humidity (\%), Dew point ($^{\circ}$C), Sea level pressure (hPa), Local pressure (hPa), Wind direction (sixteen-level scale), Vapor pressure (hPa), Precipitation (mm), Snow depth (cm), Mid-low cloud (ten-level scale), Total cloud cover (ten-level scale), Lowest cloud base (m), Solar radiation (MJ/m$^2$), Visibility (m) \\
        \midrule
        \textbf{AirKorea Observation Variables (4)}                                                                                                                                                                                                                                                                                                                                            \\
        \midrule
        Ozone (ppm), Nitrogen dioxide (ppm), Particulate matter $\le$10 $\mu$m ($\mu$g/m$^3$), Particulate matter $\le$2.5 $\mu$m ($\mu$g/m$^3$)                                                                                                                                                                                                                                               \\
        \midrule
        \textbf{Other Variables (5)}                                                                                                                                                                                                                                                                                                                                                           \\
        \midrule
        Year (2018--2021), Month (1--12), Hour (0--23), Latitude ($^{\circ}$N), Longitude ($^{\circ}$E)                                                                                                                                                                                                                                                                                        \\
        \bottomrule
    \end{tabularx}
\end{table}
In the process of merging these datasets, when multiple air quality monitoring stations were located near an ASOS observation point, we applied a method of calculating the distance based on latitude and longitude using the Haversine formula \cite{nabilla2023} and then pairing the ASOS station with the nearest air quality station. The Haversine distance approximates the great-circle distance between two points, A and B, given their latitudes and longitudes $(\phi_{A}, \phi_{B} \text{ and } \lambda_{A}, \lambda_{B})$, assuming the Earth is a perfect sphere. Typically, the Earth's radius $R\approx6371km$ is used, and the formula is expressed as equation \eqref{equation1}.
\begin{equation}
    d=2R \times \arcsin{\sqrt{\sin^2{(\frac{\phi_B-\phi_A}{2})} + \cos{\phi_A}\cos{\phi_B}\sin^2{\left(\frac{\lambda_B-\lambda_A}{2}\right)}}}
    \label{equation1}
\end{equation}
In equation \eqref{equation1}, $\phi_A$ and $\lambda_A$ represent the latitude and longitude of point A in radians, respectively. The Haversine distance d calculated using this formula represents the spherical distance (in~$km$) between the two points.

In this study, we first selected one representative meteorological observation station for each city. Then, we merged the meteorological and air quality data by pairing each station with the air quality monitoring station that had the shortest distance calculated via the Haversine formula. By forming pairs of the closest meteorological and air quality stations within each city, we were able to merge data from locations where the meteorological and air quality observation points were generally in close proximity based on latitude and longitude.

The observational data in this study contains intermittently missing values from the meteorological and air quality measurement processes, with some data points or variables partially missing at specific times. Such missing data can cause distortions in analysis results or a decrease in model prediction accuracy; therefore, we imputed them using a hierarchical approach. Specifically, short-term missing values occurring over short intervals in the time-series were handled through linear interpolation based on values from adjacent time points. For long-term missing data, such as in continuous missing intervals, data from the nearest neighboring observation station were used as substitute values. Considering the similarity of climate patterns within the same region, utilizing data from adjacent stations can minimize the potential for distortion in analysis results when imputing continuous missing data \cite{xu2022,hua2024, parra2023}. Through this procedural missing value treatment process, we were able to retain as many observations as possible for each station and time point, thereby securing the necessary data for model training. Furthermore, we generated new derived variables by applying Cyclic-Encoding \cite{porcelli2019} to the hour and month variables—which exhibit strong periodic characteristics among the time series elements—thereby enabling our models to effectively learn the temporal trends in the data.

Because meteorological and atmospheric data patterns vary annually \cite{calastrini2024}, we applied a year-based time-series split instead of simple random sampling to best reflect the temporal characteristics \cite{haris2018}. Table \ref{tab:data_folds} shows the data splitting method used in this study, which reflects the time-series characteristics. The data from 2018 to 2020 were divided using a 3-fold cross-validation approach, with each fold comprising specific years for training and validation. In the first fold, data from 2018–2019 were used for training and 2020 data for validation. In the second fold, data from 2018 and 2020 were used as the training set, and 2019 data as the validation set. In the final fold, data from 2019–2020 were used for training and 2018 data for validation, designed so that each of the three years was allocated as a validation set once. Furthermore, the 2021 data was kept as a test dataset to evaluate the final performance of the model.
\begin{table}[htbp]
    \caption{The 3-fold cross-validation data splitting strategy}
    \label{tab:data_folds}
    \begin{tabularx}{0.7\textwidth}{CCC}
        \toprule
        \textbf{Fold} & \textbf{Training Dataset} & \textbf{Validation Dataset} \\
        \midrule
        Fold 1        & 2018, 2019                & 2020                        \\
        Fold 2        & 2018, 2020                & 2019                        \\
        Fold 3        & 2019, 2020                & 2018                        \\
        \bottomrule
    \end{tabularx}
\end{table}

\subsection{Data Augmentation}\label{sec:data_augmentation}
As shown in Table \ref{tab:class_distribution}, the data for each of the six regions is concentrated in Class 2. Therefore, this section describes the SMOTENC and CTGAN models used for data augmentation of the minority classes and the data augmentation method.
\begin{table}[htbp]
    \caption{Initial distribution of visibility classes by city}
    \label{tab:class_distribution}
    \begin{tabularx}{0.75\textwidth}{@{\extracolsep\fill}ccccc}
        \toprule
        \multirow{2}{*}{Region} & \multirow{2}{*}{Sample Size} & \multicolumn{3}{c}{Ratio (\%)}                     \\
        \cmidrule(lr){3-5}
                                &                                 & Class 2                        & Class 1 & Class 0 \\
        \midrule
        Seoul                   & 35,064                          & 91.12                          & 8.73    & 0.15    \\
        Busan                   & 35,064                          & 94.54                          & 5.12    & 0.34    \\
        Incheon                 & 35,064                          & 83.46                          & 14.54   & 2.00    \\
        Daegu                   & 35,064                          & 96.34                          & 3.52    & 0.14    \\
        Daejeon                 & 35,064                          & 90.01                          & 9.35    & 0.64    \\
        Gwangju                 & 35,064                          & 90.93                          & 8.71    & 0.36    \\
        \bottomrule
    \end{tabularx}
\end{table}
SMOTE is a technique that aims to address the data imbalance problem by augmenting minority class data. This process aims to improve the stability of the training by expanding the distribution of the minority class data. However, as simple SMOTE can only learn continuous variables, we employ SMOTENC, an extended model capable of learning categorical variables as well \cite{chawla2002}.

CTGAN is a generative adversarial network (GAN)-based model developed for generating tabular data. It is designed to function effectively even with data exhibiting a mixture of categorical and continuous variables or severe class imbalance, leveraging the characteristics of GAN models that demonstrate high performance with limited data and effectively model complex probability distributions during training \cite{xu2019}.

To aim to mitigate this severe class imbalance, a Hierarchical Augmentation strategy was applied to each training data fold. This hierarchical approach was designed to allow each technique to compensate for the limitations of the other. To prevent the complex generative model CTGAN from failing to train in situations of extremely sparse data, we first augmented the data with SMOTENC and then used CTGAN for further augmentation. This was intended to generate high-quality and diverse data that could reflect not only simple linear interpolation but also the complex nonlinear distribution and interactions between variables \cite{chawla2002, temraz2025, sharma2022_gan}. To prevent information leakage, both SMOTENC and CTGAN were fit strictly within each fold's training split (i.e., two years), and all synthetic samples were added exclusively to the training data. The held-out validation year remained completely untouched by any augmentation or preprocessing learned from future data.

To verify the effectiveness of this hierarchical augmentation method, this study compared it with approaches using SMOTENC and CTGAN individually. To aim to mitigate the data imbalance issue in each of the six regions using CTGAN, we conducted experiments by individually generating 7,000, 10,000 and 20,000 data points for the minority classes, Class 0 and Class 1, respectively. That is, the 7,000 augmentation scenario means adding 7,000 data points to Class 0 and another 7,000 to Class 1, and the other scenarios were applied in the same manner. In this study, each data augmentation scenario is distinguished by naming them CTGAN7000, CTGAN10000, and CTGAN20000. Furthermore, augmentation scenarios combining SMOTENC and CTGAN are named SMOTENC\_CTGAN7000, SMOTENC\_CTGAN10000, and SMOTENC\_CTGAN20000 depending on the number of augmented data points. To ensure the reproducibility of the minority class expansion, the random state for the SMOTENC model was fixed at 42. The hyperparameter optimization for CTGAN was conducted using Optuna, an automated search framework. The search space for this process followed the configurations defined in the Supplementary Material, and the optimal hyperparameters identified through this systematic search are provided in the supplementary file (oversampling\_models\_hyperparameters.csv) available in the same repository to ensure full reproducibility.

Algorithm \ref{alg:augmentation} provides a formal pseudocode description of the fold-wise data augmentation procedure, explicitly illustrating how information leakage is prevented throughout the process.

\begin{algorithm}[htbp]
\renewcommand{\baselinestretch}{1.3}\selectfont
\caption{\textbf{Hierarchical Data Augmentation with Temporal Cross-Validation}}
\label{alg:augmentation}
\begin{algorithmic}[1]
\Require Dataset $D = \{(x_i, y_i)\}_{i=1}^{N}$ partitioned by years 2018--2020; Minority classes $\mathcal{C}_{\text{min}} = \{0, 1\}$
\Ensure Augmented training dataset $D_{\text{final}}^k$ for each fold
\For{each fold $k \in \{1, 2, 3\}$}
    \State \textbf{Partition $D$ into $D_{\text{train}}^k$ (2 years) and $D_{\text{val}}^k$ (1 held-out year)}
    \State \Comment{Temporal split ensures no future information leakage}
    \State \textbf{Extract minority subset:} $D_{\text{min}}^k \gets \{(x,y) \in D_{\text{train}}^k : y \in \mathcal{C}_{\text{min}}\}$
    \State \textbf{Fit SMOTENC exclusively on $D_{\text{train}}^k$} \Comment{No access to $D_{\text{val}}^k$}
    \State Generate synthetic minority samples $D_{\text{syn}}^{\text{SMOTE}}$ for classes in $\mathcal{C}_{\text{min}}$
    \State $D_{\text{aug}}^k \gets D_{\text{train}}^k \cup D_{\text{syn}}^{\text{SMOTE}}$
    \State \textbf{Fit CTGAN exclusively on $D_{\text{aug}}^k$} \Comment{Hierarchical augmentation}
    \State Generate synthetic minority samples $D_{\text{syn}}^{\text{CTGAN}}$ for classes in $\mathcal{C}_{\text{min}}$
    \State $D_{\text{final}}^k \gets D_{\text{aug}}^k \cup D_{\text{syn}}^{\text{CTGAN}}$
\EndFor
\State \Return Augmented datasets $\{D_{\text{final}}^k\}_{k=1}^{3}$ for model training
\end{algorithmic}
\end{algorithm}

\subsection{Two-Stage Soft Voting Ensemble}\label{sec:two_stage_soft_voting}

\begin{figure}[htbp]
    \centering
    \includegraphics[width=0.65\textwidth]{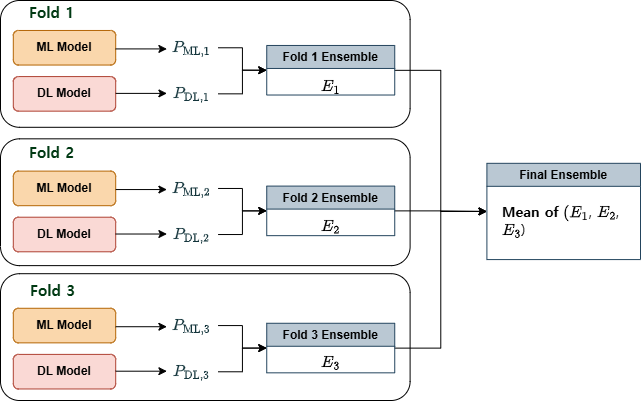}
    \caption{Diagram of the model two stage ensemble method}\label{fig:ensemble_diagram}
\end{figure}

To address prediction stability and performance, we implemented a two-stage soft voting ensemble structure that sequentially combines predictions at both the model level and the fold level, as shown in figure \ref{fig:ensemble_diagram}. This approach aims to leverage the complementary strengths of different model architectures and mitigate overfitting to specific temporal subsets of the training data.

\subsubsection{Model Stage Ensemble (Within-Fold Fusion)}
Within each fold $k$ ($k=1, 2, 3$) of the 3-fold cross-validation, we identify the best-performing machine learning (ML) and deep learning (DL) models for each city. These two models are then combined via soft voting to produce a single fold-specific ensemble model $E_k$. Soft voting determines the final predicted class by taking the arithmetic mean of the class-wise probabilities predicted by each model \cite{abdullah2025, cao2015, caovan2024, khan2023, kibria2022, manconi2022}. As illustrated in Figure \ref{fig:ensemble_diagram}, for a given input sample, each of the two models ($N=2$) outputs a probability vector containing the prediction probabilities for classes 0, 1, and 2 (denoted as $P_{\text{ML}}$ for the ML model and $P_{\text{DL}}$ for the DL model). The ensemble probability vector ($\bar{P}$) for fold $k$ is calculated as shown in equation \eqref{equation2}:
\begin{equation}
    \bar{P}=\frac{1}{N}\sum_{i=1}^{N}P_i = \frac{P_{\text{ML}} + P_{\text{DL}}}{2}
    \label{equation2}
\end{equation}
The $\bar{P}$ value obtained via the method described in equation~\eqref{equation2} is used as the probability vector $E_k$ for that fold. The probability vectors from all three folds are then subjected to soft voting once more, returning the final probability vector from the double soft voting process.

Since this process is performed independently for each fold, the resulting three ensemble models $E_1, E_2,$ and $E_3$ produce robust prediction results tailored to the data characteristics of their respective training-validation splits.

\subsubsection{Fold Stage Ensemble (Cross-Fold Aggregation)}

Through the 3-fold cross-validation, three independent ensemble models—$E_1, E_2$, and $E_3$—are generated, each trained on different year combinations. To overcome the limitations of a single model's potential overfitting to specific temporal subsets and to maximize generalization performance, we employ a second stage of soft voting that aggregates the predictions from these three fold-specific ensembles \cite{sultan2025, sharma2025_fault}. For a given test sample $X_{\text{test}}$ from the 2021 test dataset, each of the three fold-specific ensemble models outputs a class prediction probability vector:
\begin{itemize}
    \item Prediction probability of the Fold 1 ensemble model $(E_1): P_1 = E_1(X_{\text{test}}) = [p_{1,0}, p_{1,1}, p_{1,2}]$

    \item Prediction probability of the Fold 2 ensemble model $(E_2): P_2 = E_2(X_{\text{test}}) = [p_{2,0}, p_{2,1}, p_{2,2}]$

    \item Prediction probability of the Fold 3 ensemble model $(E_3): P_3 = E_3(X_{\text{test}}) = [p_{3,0}, p_{3,1}, p_{3,2}]$
\end{itemize}

The three prediction probability vectors obtained in this way are arithmetically averaged as shown in equation \eqref{equation3} to calculate the final prediction probability vector.
\begin{equation}
    P_{\text{final}}=\frac{1}{3}\sum_{k=1}^{3}P_k=\left[\frac{p_{1,0}+p_{2,0}+p_{3,0}}{3},\ \frac{p_{1,1}+p_{2,1}+p_{3,1}}{3},\frac{p_{1,2}+p_{2,2}+p_{3,2}}{3}\right]
    \label{equation3}
\end{equation}

During this process, the year data corresponding to the validation set for each fold was all sourced from the original data, thereby preventing data leakage caused by augmented data in advance. This approach has the following theoretical basis. First, since each fold-specific ensemble model is trained on data combinations from different years, they learn different perspectives on the temporal variation of the data distribution. By averaging their predictions, we aim to achieve a robust prediction that is not confined to a specific year or season. Second, it is intended to increase prediction stability by canceling out the variance contained in the prediction errors of individual models \cite{sultan2025}.

Finally, the class with the highest value in the calculated probability vector $P_{\mathrm{final}}$ is determined as the final predicted class ($\hat{y}$), as shown in equation \eqref{equation4}.
\begin{equation}
    \hat{y} = \operatornamewithlimits{argmax}_{c \in \{0, 1, 2\}} (P_{\text{final}})
    \label{equation4}
\end{equation}

This two-stage soft voting ensemble structure complements the prediction limitations of a single model by combining predictions at both the model level and the cross-validation fold level, aiming to address the problems of data imbalance and distribution shift \cite{sultan2025, sharma2025_fault}.

\subsection{Evaluation Metrics}\label{sec:evaluation_metrics}

In this study, we used the multiclass Critical Success Index ($\mathrm{CSI}$), Matthews Correlation Coefficient ($\mathrm{MCC}$), and Accuracy ($\mathrm{ACC}$) as metrics for model performance evaluation.
First, $\mathrm{CSI}$ is a widely used metric for evaluating the detection capability of minority classes. In Table \ref{tab:confusion_matrix_metrics}, $H_{ij}, M_{ij},$ and $F_{ij}$ denote the number of samples whose true class is $i$ and that are predicted as class $j$. For example, $M_{02}$ refers to cases where an actual Class 0 instance is incorrectly predicted as Class 2. Since this study focuses on detecting low-visibility events, the components of the evaluation metrics—H, M, and F—are defined centered on Class 0 and Class 1 as follows:

The following formulae are to be used for each entry:
\begin{itemize}
    \item H (Hit): The number of cases where actual Class 0 or 1 was correctly predicted.
          \begin{equation}
              H = H_{00} + H_{11}
              \label{equation5}
          \end{equation}

    \item M (Miss): The number of cases where actual Class 0 or 1 was incorrectly predicted as Class 2.
          \begin{equation}
              M = M_{02} + M_{12}
              \label{equation6}
          \end{equation}

    \item F (False Alarm): The number of cases where actual Class 2 was incorrectly predicted as Class 0 or 1.
          \begin{equation}
              F = F_{01} + F_{10} + F_{20} + F_{21}
              \label{equation7}
          \end{equation}
\end{itemize}
\begin{table}[htbp]
    \centering
    \caption{The multiclass confusion matrix for performance evaluation}
    \label{tab:confusion_matrix_metrics}

    \begin{tabularx}{0.7\textwidth}{lCCCC}
        \toprule
        \multicolumn{2}{c}{}          & \multicolumn{3}{c}{Predicted Class}                                                                                                          \\
        \cmidrule(lr){3-5}
                                      &                                     & Class 0                          & Class 1                          & Class 2                          \\
        \midrule
        \multirow{3}{*}{Actual Class} & Class 0                             & \cellcolor[HTML]{32CB00}$H_{00}$ & \cellcolor[HTML]{FD6864}$F_{01}$ & \cellcolor[HTML]{FFC702}$M_{02}$ \\
                                      & Class 1                             & \cellcolor[HTML]{FD6864}$F_{10}$ & \cellcolor[HTML]{32CB00}$H_{11}$ & \cellcolor[HTML]{FFC702}$M_{12}$ \\
                                      & Class 2                             & \cellcolor[HTML]{FD6864}$F_{20}$ & \cellcolor[HTML]{FD6864}$F_{21}$ & $C_{22}$                         \\
        \bottomrule
    \end{tabularx}
\end{table}

$\mathrm{CSI}$ is a widely used metric for evaluating the detection capability of minority classes and is defined as in equation \eqref{equation8}.
\begin{equation}
    CSI=\frac{H}{H+M+F}=\frac{H_{00}+H_{11}}{H_{00}+H_{11}+M_{02}+M_{12}+F_{01}+F_{10}+F_{20}+F_{21}}
    \label{equation8}
\end{equation}

The $\mathrm{CSI}$ calculated in this manner has a value between 0 and 1, where a value closer to 1 indicates better prediction for the minority classes. $\mathrm{CSI}$ allows for a precise evaluation of how effectively the model detects each of the minority classes, Class 0 and Class 1, not just its overall accuracy \cite{joseph1990}.

$\mathrm{MCC}$ is a metric that represents the correlation between the predicted and actual values on a scale from -1 to 1. In a multiclass situation, it can be calculated as a Generalized $\mathrm{MCC}$ using the entire confusion matrix. This Generalized MCC reflects all TP (True Positive), TN (True Negative), FP (False Positive), and FN (False Negative) values from the $K\times K$ confusion matrix at once, providing a single correlation coefficient that indicates how balanced the predictions for each class are. A value close to 1 indicates a strong positive correlation between prediction and reality, 0 indicates a random level, and a value close to -1 indicates a negative correlation. Since $\mathrm{MCC}$ is useful for evaluating the overall prediction balance of a model even in a class-imbalanced environment \cite{imani2025}, this study used it alongside $\mathrm{CSI}$ as a primary metric to check not only the detection of poor visibility but also the prediction stability for the remaining classes.

The final metric used, $\mathrm{ACC}$, is a measure that indicates the proportion of correct predictions made by the model out of the total number of samples. In a multiclass problem, it is calculated by dividing the total number of correct predictions across all classes in the confusion matrix by the total number of samples. This value ranges from 0 to 1, with a value closer to 1 indicating a more accurate model. Although a high $\mathrm{ACC}$ can be misleading when poor visibility cases are extremely rare, as the model might miss low-visibility instances, this study used $\mathrm{ACC}$ as a supplementary metric and designated $\mathrm{CSI}$ and $\mathrm{MCC}$ as the primary metrics for evaluation.

\subsection{Distributional Distance Metric}\label{sec:distributional_distance_metric}

To quantify distributional similarity between datasets---both for evaluating synthetic data fidelity and for measuring distribution shifts between training and test periods---this study employs the Wasserstein distance. The Wasserstein distance is an intuitive metric for measuring distance between distributions, accounting for changes in shape. It represents the minimum cost required to move the mass constituting the two distributions from one location to another. Mathematically, if the data distributions for two groups are denoted as $\mu$ and $\nu$, respectively, the $p$-Dimension Wasserstein distance $W_{p}(\mu, \nu)$ is defined as in equation \eqref{equation9} \cite{martin2017}.

\begin{equation}
    W_{p}(\mu, \nu) = \left( \inf_{\gamma \in \Gamma(\mu, \nu)} \int_{\mathbb{R}^p \times \mathbb{R}^p} \|x-y\|^p \mathrm{d}\gamma(x,y) \right)^{\frac{1}{p}}
    \label{equation9}
\end{equation}

where $\Gamma(\mu,\nu)$ denotes the set of all joint distributions with marginal distributions $\mu$ and $\nu$.

\section{Prediction Models}\label{sec:prediction_models}

In each fold of Table \ref{tab:data_folds}, a multiclass visibility prediction model was implemented using five models: the machine learning models XGBoost and LightGBM, and the deep learning models ResNet-like, FT-Transformer, and DeepGBM. The characteristics and rationale for selecting each model are described below.

\begin{description}
    \item[eXtreme Gradient Boosting (XGBoost)] ensures fast training speed even with large-scale data through the Gradient Boosting method. This allows it to achieve stable and high prediction performance even in problems like visibility nowcasting, which involve complex interactions of various meteorological and air quality variables or imbalanced data \cite{chen2016}.
    
    \item[Light Gradient Boosting Machine (LightGBM)] is also a Gradient Boosting algorithm, similar to XGBoost. However, LightGBM offers faster training speed and higher efficiency on large datasets compared to XGBoost \cite{ke2017}, making it well-suited for visibility nowcasting problems that involve complex meteorological and air quality data.
    
    \item[Residual Network like (ResNet-like)] models leverage skip connections (residual blocks) that enable training of deeper networks by mitigating the vanishing gradient problem \cite{he2016}. In this study, considering the characteristics of meteorological data, we embraced this residual block-based architecture that is deeper, more expressive, and more stable.
    
    \item[Feature Tokenizer+Transformer (FT-Transformer)] is a model that applies the Transformer architecture, primarily used in natural language processing, to tabular data. It effectively learns complex and non-linear global dependencies between features by utilizing the self-attention mechanism \cite{gorishniy2021}. Therefore, we adopt this architecture, which learns complex inter-variable characteristics in a short time and effectively identifies changes over time.
    
    \item[Deep Gradient Boosting Machine (DeepGBM)] is a hybrid model that combines the strengths of the decision tree-based GBDT (Gradient Boosted Decision Tree) and a Deep Neural Network (DNN). In this study, the CatNN module was applied for processing categorical variables, and the GBDT2NN module was used for numerical variables. CatNN is responsible for DNN-based processing of categorical data, while GBDT2NN is designed to transfer information learned from GBDT to the neural network. This allows for the effective reflection of the diverse characteristics of meteorological data \cite{ke2019}.
\end{description}

To ensure the reproducibility of our experiments and address the technical complexity of the five predictive models, the hyperparameter optimization methodology and the specific search spaces are fully documented in the Supplementary Material. Furthermore, the final optimal hyperparameters identified through this systematic optimization process are organized and provided for each model within the \texttt{best\_params\_per\_models} directory of the same repository.

\section{Empirical Analysis}\label{sec:empirical_analysis}

This section evaluates the proposed methodology through three stages: (1) assessing the fidelity of augmented data to the original distribution, (2) analyzing the impact of data augmentation and ensemble techniques on model performance, and (3) investigating the causes of performance degradation observed in test data through distribution shift analysis.

\subsection{Synthetic Data Quality Assessment}\label{sec:synthetic_data_quality}

\begin{figure}[H]
    \centering
    \begin{minipage}{0.42\textwidth}
        \centering
        \includegraphics[width=\linewidth]{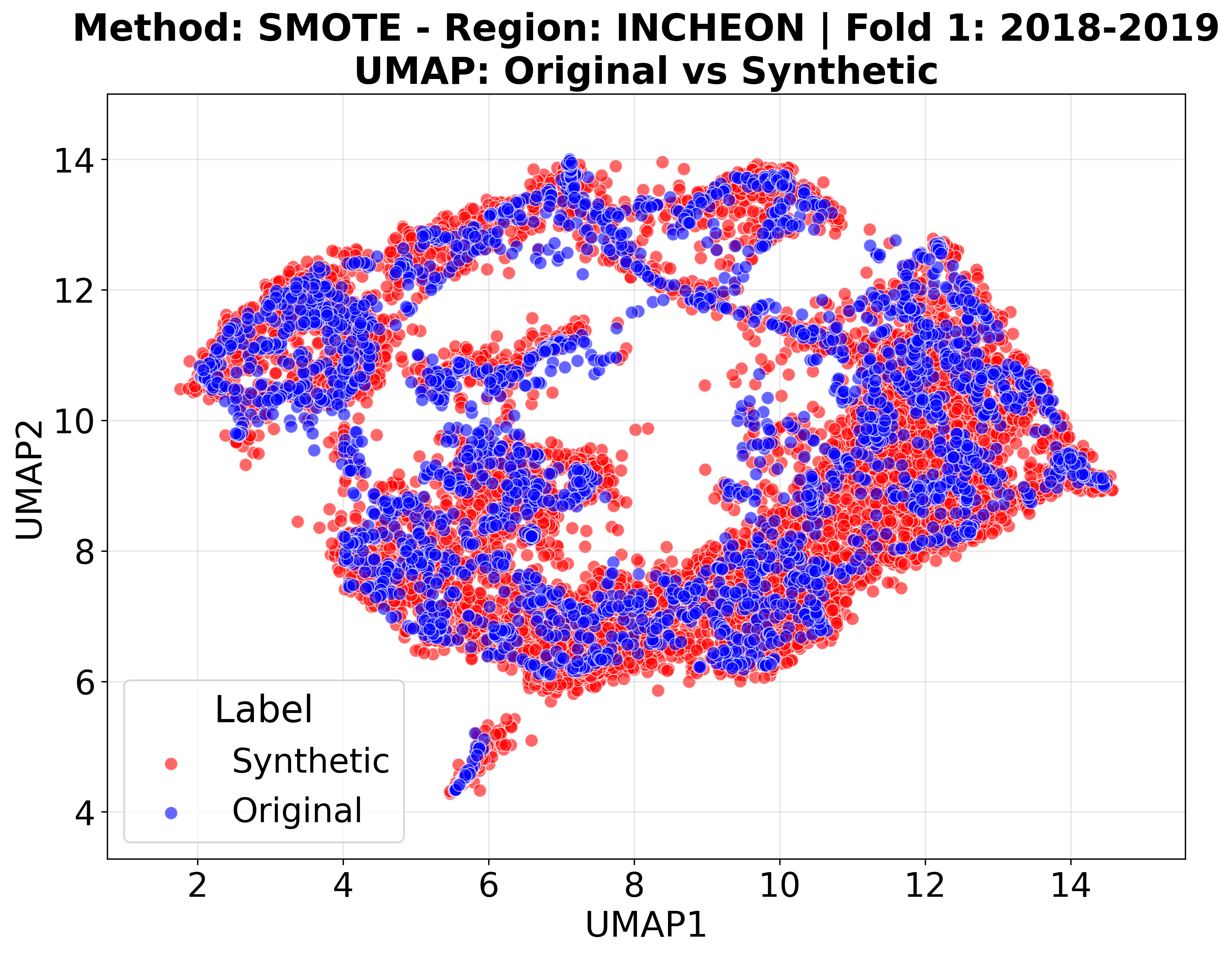}
    \end{minipage}
    \hfill
    \begin{minipage}{0.48\textwidth}
        \centering
        \includegraphics[width=\linewidth]{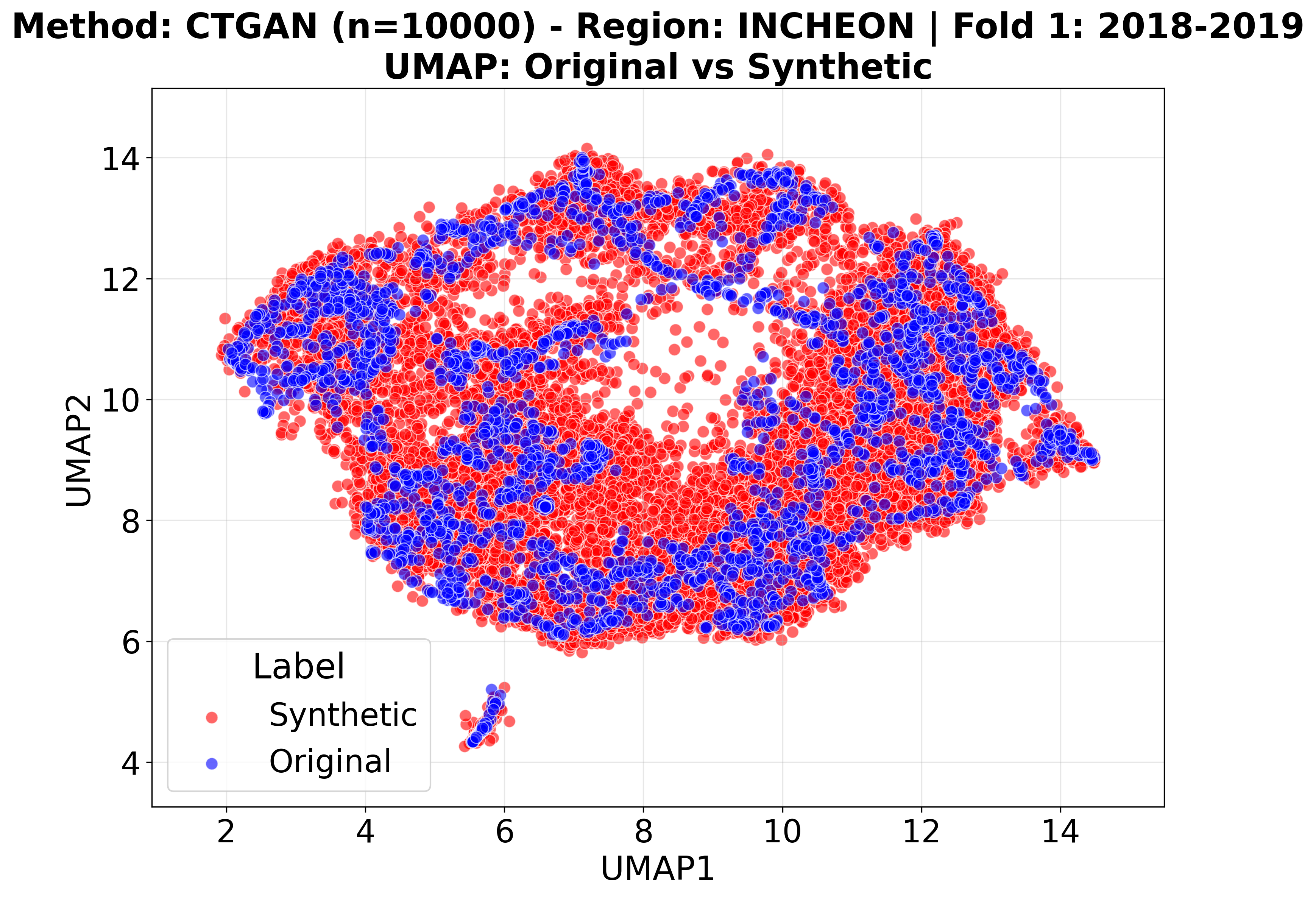}
    \end{minipage}
    \par\medskip
    \begin{minipage}{0.53\textwidth}
        \centering
        \includegraphics[width=\linewidth]{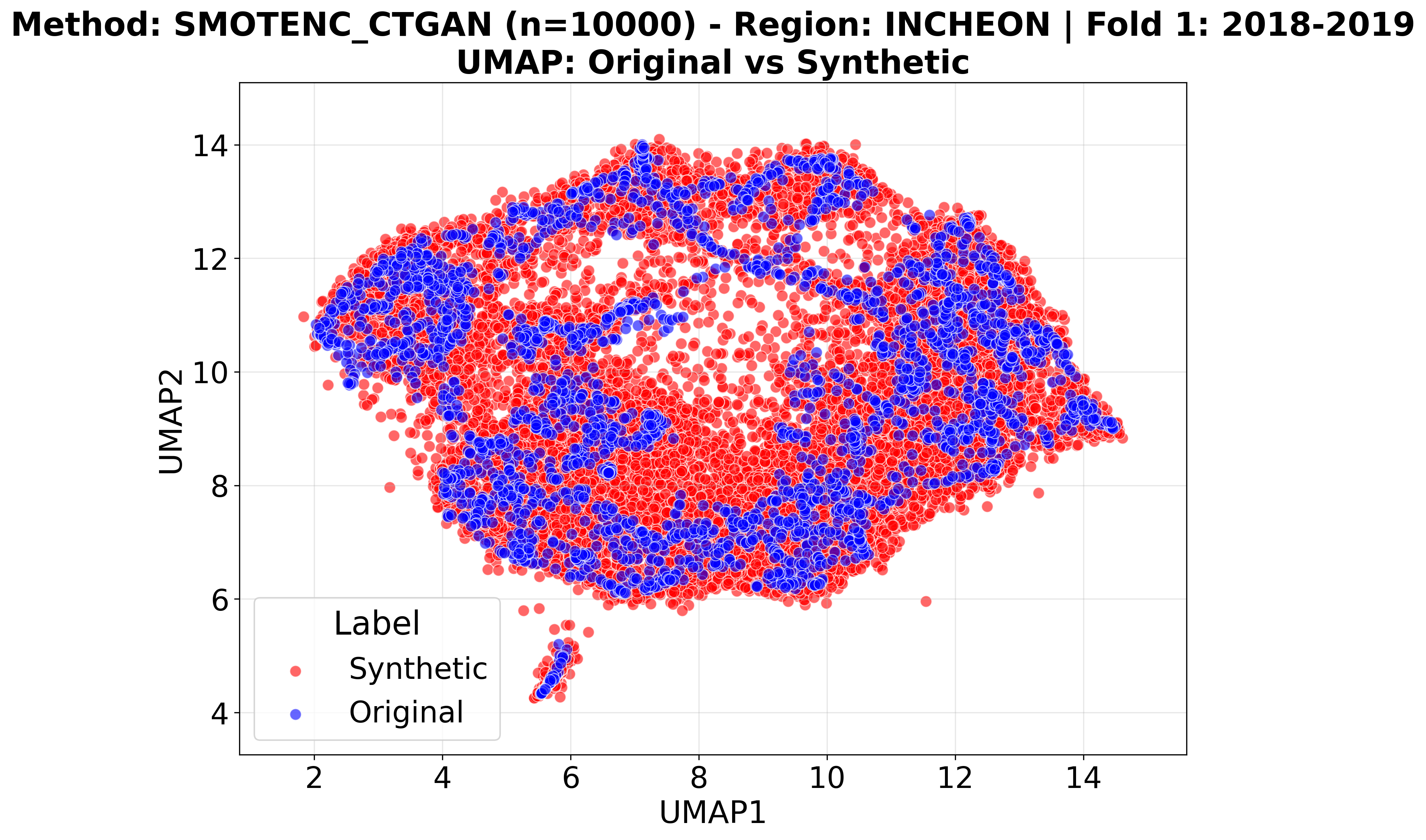}
    \end{minipage}
    \caption{UMAP visualization comparing the distributions of original and augmented data for Incheon (Fold 1). Each subplot shows a different augmentation method: SMOTENC (left), CTGAN10000 (right), and SMOTENC\_CTGAN10000 (bottom).}
    \label{fig:umap_comparison}
\end{figure}

The most fundamental requirement for augmented data is not so much the improvement of predictive model performance, but rather how well it follows the distribution of the original data. Therefore, in this study, we plotted UMAP to visually verify how well each augmented data point followed the distribution of the original data \cite{tian2024}. To ensure methodological rigor and reproducibility in the UMAP visualization, we applied a strict protocol: the StandardScaler and UMAP model (fit\_transform) were fitted exclusively on the original training data to learn the scaling parameters and low-dimensional manifold structure, respectively. Subsequently, all augmented data were transformed using only the learned scaler and projected onto the learned manifold via the transform method, thereby preventing data leakage and ensuring that synthetic data did not influence the reference distribution. The UMAP random state was fixed at 42, with hyperparameters n\_neighbors set to 30 and min\_dist set to 0.1 for full reproducibility. Figure \ref{fig:umap_comparison} shows representative UMAP plots for Incheon (Fold 1) depicting the distribution of augmented data relative to the original training data. The full visualization graphs for all regions, all folds, and all augmentation methods are provided in the \texttt{umap\_plots} directory of the Supplementary Material repository.

While UMAP provides useful visual confirmation of overall distributional similarity, it does not capture variable-specific fidelity. To rigorously assess how well the augmented data preserves the marginal distributions of individual variables, we employed the Wasserstein Distance (WD) as defined in Section \ref{sec:distributional_distance_metric}, along with Kernel Density Estimation (KDE) plots for visual inspection.

Table \ref{tab:wd_fidelity} presents WD for Incheon (Fold 1), focusing on Relative Humidity (RH) and PM2.5---two variables widely recognized as primary determinants of atmospheric visibility in meteorological studies \cite{won2020, sun2020}. Following the same protocol used for UMAP, the original data was used to fit a StandardScaler, and all augmented data were transformed using this learned scaler to ensure fair comparison.
\clearpage
\begin{table}[htbp]
    \centering
    \caption{Wasserstein distance for RH and PM2.5 variables in Incheon (Fold 1), evaluating synthetic data fidelity across augmentation methods.}
    \label{tab:wd_fidelity}
    \begin{tabular}{@{}cccc@{}}
        \toprule
        Class                    & Variable                 & Method       & \begin{tabular}[c]{@{}c@{}}Wasserstein\\ Distance\end{tabular} \\
        \midrule
        \multirow{6}{*}{Class 0} & \multirow{3}{*}{RH (\%)} & SMOTENC      & \textbf{0.0570}                            \\
                                 &                          & CTGAN10000   & 0.3546                                       \\
                                 &                          & Hybrid\tnote{a} & 0.4965                                       \\
        \cmidrule(lr){2-4}
                                 & \multirow{3}{*}{PM2.5}   & SMOTENC      & \textbf{0.0252}                            \\
                                 &                          & CTGAN10000   & 0.1124                                    \\
                                 &                          & Hybrid\tnote{a} & 0.2657                                    \\
        \midrule
        \multirow{6}{*}{Class 1} & \multirow{3}{*}{RH (\%)} & SMOTENC      & \textbf{0.0294}                            \\
                                 &                          & CTGAN10000   & 0.1458                                    \\
                                 &                          & Hybrid\tnote{a} & 0.2224                                    \\
        \cmidrule(lr){2-4}
                                 & \multirow{3}{*}{PM2.5}   & SMOTENC      & \textbf{0.0029}                            \\
                                 &                          & CTGAN10000   & 0.0898                                     \\
                                 &                          & Hybrid\tnote{a} & 0.0691                                     \\
        \bottomrule
    \end{tabular}
    \begin{tablenotes}
        \item[a] SMOTENC\_CTGAN10000.
    \end{tablenotes}
\end{table}

\begin{figure}[htbp]
    \centering
    \includegraphics[width=1\textwidth]{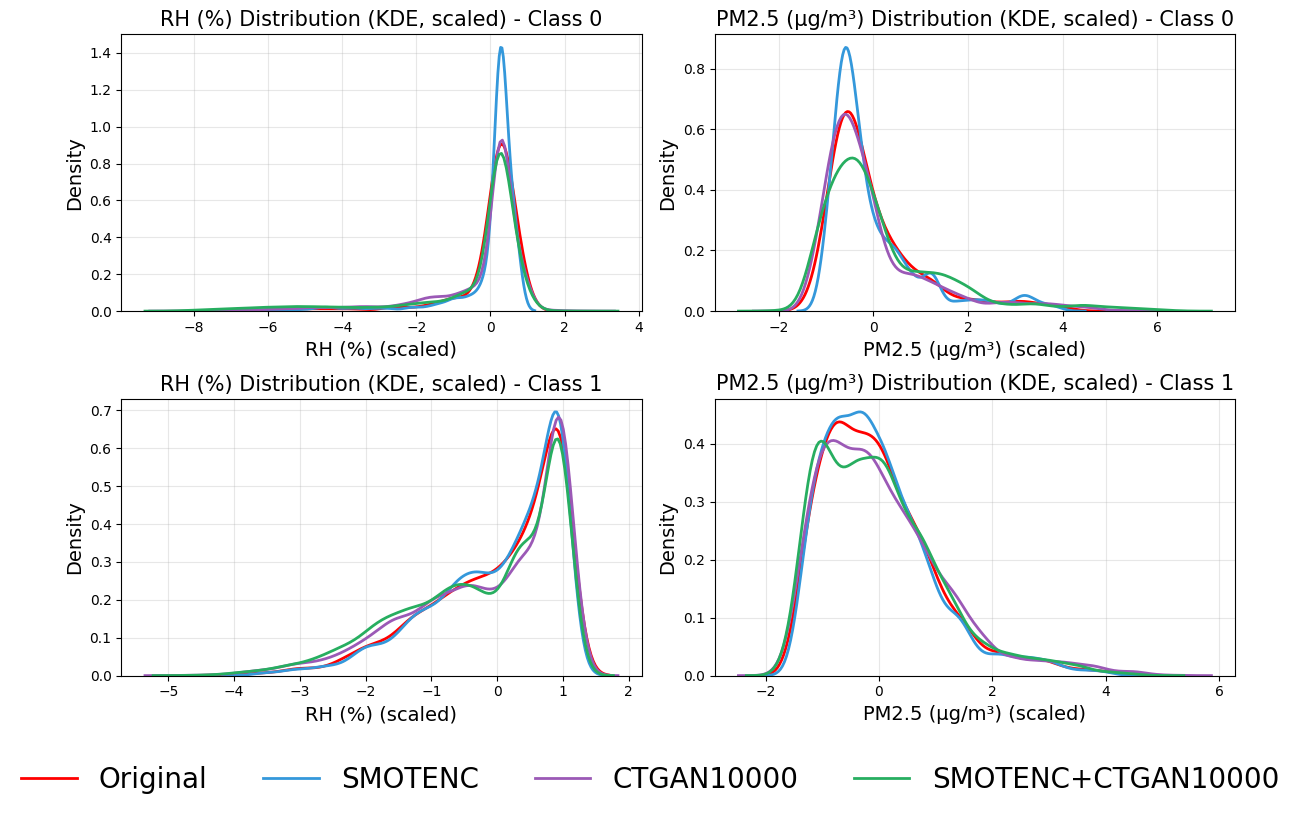}
    \caption{KDE plots for RH and PM2.5 comparing original and augmented data distributions (SMOTENC, CTGAN10000, and Hybrid) for Incheon (Fold 1).}
    \label{fig:kde_incheon}
\end{figure}

Table \ref{tab:wd_fidelity} reveals that SMOTENC consistently achieves the lowest WD values across both classes and variables, indicating high fidelity to the original marginal distributions. In contrast, CTGAN and Hybrid methods exhibit higher WD values, particularly for RH in Class 0. However, lower WD does not imply that SMOTENC universally preserves the original distribution better. As a linear interpolation method, SMOTENC tends to concentrate samples in high-density regions, which can lead to artificially peaked distributions. This effect is visible in the KDE plots (Figure \ref{fig:kde_incheon}): SMOTENC (blue) exhibits noticeably higher kurtosis than the original distribution (red), whereas CTGAN10000 (purple) show smoother, more realistic shapes despite its higher WD values.
Given these trade-offs between distributional fidelity and sample diversity, we complement the distributional analysis with downstream classification performance evaluation in the following sections. While both Wasserstein distance and kernal density estimation provide direct evidence of how well each method preserves the statistical properties of the original data, downstream performance offers an additional perspective on their practical utility for predictive modeling. We systematically evaluate how each augmentation method affects model performance across multiple architectures, recognizing that the optimal choice depends on both the distributional characteristics documented here and the downstream performance.

\subsection{Model Training and Hyperparameter Optimization}\label{sec:model_training_optimization}

Exploring hyperparameters for all permutations across annual 3-fold cross-validation, six regions, seven augmented datasets, and five model combinations would be computationally prohibitive. Therefore, this study adopted a stepwise optimization strategy with an initial screening phase. This strategy aims to effectively reduce the overall search space by first utilizing relatively lightweight tree-based models to select promising augmentation scenarios, then performing sophisticated hyperparameter optimization for deep learning models only on the selected minority of scenarios.

First, an initial screening was conducted using tree-based boosting models such as XGBoost and LightGBM with their default hyperparameters, performing year-based 3-fold cross-validation for each region and each augmented dataset (including the original data). Table \ref{tab:model_rank_counts} lists the counts of the best rank cases for each dataset, and the average performance across folds in sequential order. At this stage, we sought to select one augmentation dataset with the highest validation performance from each of the following three categories (SMOTENC, CTGAN category, SMOTENC\_CTGAN hybrid category) for each domain. Here, for the CTGAN model, the primary metric “Number of times selected as Best performance” was identical across all categories, and the average CSI performance was also measured as highly similar. At this point, for the SMOTENC\_CTGAN model, SMOTENC\_CTGAN10000 demonstrated the most outstanding performance across both metrics we measured, and was therefore selected. Consequently, when comparing the performance of augmented data, it was deemed advantageous to maintain the same augmented data size for both CTGAN and SMOTENC\_CTGAN. Thus, the CTGAN model also adopted CTGAN10000.

\begin{table}[htbp]
    \centering
    \caption{Screening results for representative augmentation samples within each strategy category based on pairwise competition}
    \label{tab:model_rank_counts}
    \begin{tabular}{llcccl}
        \toprule
        \textbf{Category}             & \textbf{Data Sample}                 & \textbf{Best Rank Cases}\tnote{a} & \textbf{Mean CSI} & \textbf{Selection} \\ \midrule
        \multirow{3}{*}{CTGAN-only}   & CTGAN7000           & 4                                 & 0.4958            & -                  \\
                                      & CTGAN10000          & 4                                 & 0.4955            & \textbf{Selected}  \\
                                      & CTGAN20000          & 4                                 & 0.4968            & -                  \\ \midrule
        \multirow{3}{*}{Hybrid (S+C)} & SMOTENC\_CTGAN7000  & 1                                 & 0.4881            & -                  \\
                                      & SMOTENC\_CTGAN10000 & \textbf{7}                        & 0.4905            & \textbf{Selected}  \\
                                      & SMOTENC\_CTGAN20000 & 4                                 & 0.4902            & -                  \\ \midrule
        Baseline                      & pure                                 & -                                 & 0.4782            & \textbf{Selected}  \\
        Linear                        & SMOTENC             & \textbf{10}                       & 0.5201            & \textbf{Selected}  \\ \bottomrule
    \end{tabular}
    \begin{tablenotes}
        \item[a] Number of times the sample ranked first within its category across all region-fold combinations using unoptimized XGBoost and LightGBM.
    \end{tablenotes}
\end{table}

\begin{figure}[H]
    \centering
    \includegraphics[width=1\textwidth]{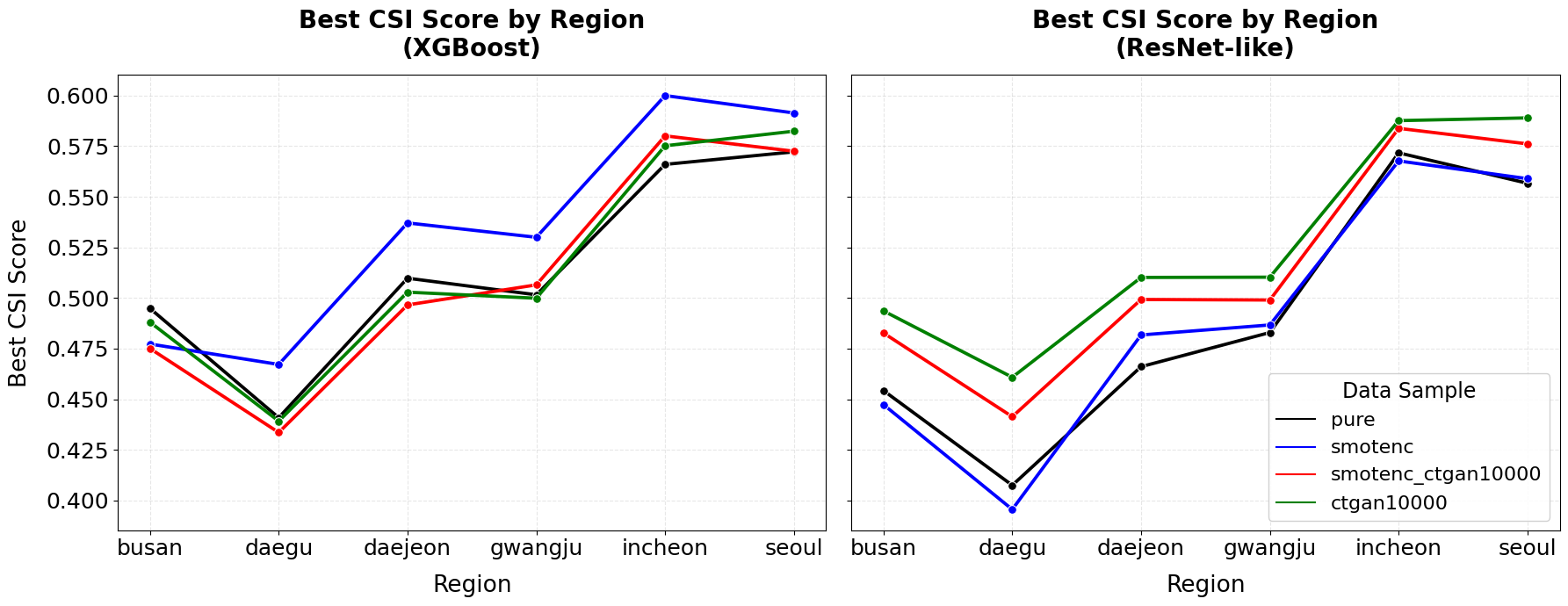}
    \caption{$\mathrm{CSI}$ performance improvement from data augmentation, categorized by model and city.}
    \label{fig:csi_improvement}
\end{figure}

We proceed to compare the validation performance of all prediction models across regions and data types, using the three selected augmented datasets and the original data as shown in Table \ref{tab:model_rank_counts}. Figure \ref{fig:csi_improvement} shows the results of the best augmentation combination for each of the 6 regions and 5 models, and the performance change before and after augmentation. It demonstrates the impact of data augmentation techniques on the performance of each model, revealing that the effect varies dramatically depending on the model architecture. The most noteworthy finding is that, within deep learning-based models, single CTGAN augmentation yielded the highest performance, albeit marginally. The CTGAN dataset demonstrated performance that was either very similar to or slightly superior to that of the SMOTENC\_CTGAN dataset across all regions.
This outcome complements the WD and KDE metrics, which previously yielded conflicting results, enabling the selection of optimal augmentation data for each learning model. CTGAN10000 was identified as the most effective augmentation data for deep learning models, while SMOTENC proved most suitable for machine learning models.

For detailed best $\mathrm{CSI}$ scores by augmentation method for each region, see Appendix \ref{app:best_csi_by_model}.

\subsection{Ensemble Performance Evaluation}\label{sec:ensemble_evaluation}

To further increase the prediction stability and accuracy using the individual models whose performance was improved through data augmentation, we performed the two-stage soft voting mentioned in Section \ref{sec:two_stage_soft_voting}. For each region, we combined the highest-performing single model from both the ML and DL models to explore the optimal ensemble \cite{sfar2025}, with the results summarized in Table \ref{tab:best_ensemble_results}.

Table \ref{tab:best_ensemble_results} shows the best performing ensemble combination for each region and the corresponding $\mathrm{CSI}$ scores. The analysis revealed that the ensemble technique showed the synergistic effect in Incheon(+0.018) and Seoul(+0.013), outperforming the single models. In contrast, a slight decrease in performance was observed in Gwangju(-0.001) and Daejeon(-0.008).
\begin{table}[htbp]
    \centering
    \footnotesize
    \caption{Validation performance of the optimal two-stage soft voting ensemble combination for each city}
    \label{tab:best_ensemble_results}
    \begin{tabular}{lcccccc}
        \toprule
        Region  & \begin{tabular}[c]{@{}c@{}}Best ML Model\\ (Data Sample)\end{tabular}        & CSI(ML) & \begin{tabular}[c]{@{}c@{}}Best DL Model\\ (Data Sample)\end{tabular} & CSI(DL) & Ensemble CSI\tnote{a} & Improvement\tnote{b} \\ \midrule
        Seoul   & \begin{tabular}[c]{@{}c@{}}XGBoost\\ (SMOTENC)\end{tabular} & 0.5914  & \begin{tabular}[c]{@{}c@{}}FT-Transformer\\ (CTGAN10000)\end{tabular} & 0.5937  & 0.6071                & +0.0134              \\
        Busan   & \begin{tabular}[c]{@{}c@{}}XGBoost\\ (pure)\end{tabular}                     & 0.4949  & \begin{tabular}[c]{@{}c@{}}FT-Transformer\\ (CTGAN10000)\end{tabular} & 0.4960  & 0.4982                & +0.0022              \\
        Incheon & \begin{tabular}[c]{@{}c@{}}XGBoost\\ (SMOTENC)\end{tabular} & 0.6000  & \begin{tabular}[c]{@{}c@{}}ResNet-like\\ (CTGAN10000)\end{tabular}    & 0.5876  & 0.6018                & +0.0180              \\
        Daegu   & \begin{tabular}[c]{@{}c@{}}XGBoost\\ (SMOTENC)\end{tabular} & 0.4672  & \begin{tabular}[c]{@{}c@{}}ResNet-like\\ (CTGAN10000)\end{tabular}    & 0.4609  & 0.4719                & +0.0047              \\
        Daejeon & \begin{tabular}[c]{@{}c@{}}XGBoost\\ (SMOTENC)\end{tabular} & 0.5371  & \begin{tabular}[c]{@{}c@{}}ResNet-like\\ (CTGAN10000)\end{tabular}    & 0.5102  & 0.5294                & -0.0077              \\
        Gwangju & \begin{tabular}[c]{@{}c@{}}XGBoost\\ (SMOTENC)\end{tabular} & 0.5300  & \begin{tabular}[c]{@{}c@{}}DeepGBM\\ (CTGAN10000)\end{tabular}        & 0.5204  & 0.5289                & -0.0011              \\ \bottomrule
    \end{tabular}
    \begin{tablenotes}
        \item[a] Validation $\mathrm{CSI}$ performance with two-stage soft voting using the best ML and DL models.
        \item[b] $CSI_{\mathrm{Ensemble}} - \max(CSI_{\mathrm{ML}}, CSI_{\mathrm{DL}})$
    \end{tablenotes}
\end{table}

\subsection{Test Performance and Distribution Shift Analysis}\label{sec:test_performance_analysis}

Table \ref{tab:final_test_performance} shows the final model performance test using the test dataset, compared to the performance of the validation dataset for the combination of the best-performing ensemble by region from Table \ref{tab:best_ensemble_results}.
As seen in Table \ref{tab:final_test_performance}, performance on the test dataset changed across all regions compared to the validation dataset. While Busan, Incheon, and Gwangju showed modest performance degradation of 2-7\%, Seoul, Daegu, and Daejeon exhibited substantial drops of 23-42\%.

\begin{table}[htbp]
    \centering
    \caption{Performance comparison between validation and test datasets}
    \label{tab:final_test_performance}
    \begin{tabular}{@{}lccccc@{}}
        \toprule
        Region  & \begin{tabular}[c]{@{}c@{}}Validation\\ CSI (A)\end{tabular} & \begin{tabular}[c]{@{}c@{}}Test\\ CSI (B)\end{tabular} & \begin{tabular}[c]{@{}c@{}}Performance\\ Gap ($\Delta$)\tnote{a}\end{tabular} & \begin{tabular}[c]{@{}c@{}}Change\\ (\%)\tnote{b}\end{tabular} \\ \midrule
        Seoul   & 0.6071                                                       & 0.3521                                                 & -0.2550                                                                       & -42.0                                                          \\
        Busan   & 0.4982                                                       & 0.4712                                                 & -0.0270                                                                       & -5.4                                                           \\
        Incheon & 0.6018                                                       & 0.5615                                                 & -0.0403                                                                       & -6.7                                                           \\
        Daegu   & 0.4719                                                       & 0.2739                                                 & -0.1980                                                                       & -42.0                                                          \\
        Daejeon & 0.5294                                                       & 0.4078                                                 & -0.1216                                                                       & -23.0                                                          \\
        Gwangju & 0.5289                                                       & 0.5052                                                 & -0.0237                                                                       & -4.5                                                           \\ \bottomrule
    \end{tabular}
    \begin{tablenotes}
        \item[a] $\Delta = B - A$
        \item[b] $\mathit{Change}(\%) = \dfrac{\Delta}{A} \times 100$
    \end{tablenotes}
\end{table}

To investigate the cause of this performance degradation, this study analyzed the confusion matrices of the test dataset (see Appendix \ref{app:confusion_matrices} for the full confusion matrices by region). The analysis revealed a notable pattern: in the three regions with substantial performance drops (Seoul, Daegu, Daejeon), the model frequently misclassified Class 2 (normal visibility) samples as Class 1 (reduced visibility) and vice versa. To quantify this phenomenon, we computed a binary classification $\mathrm{CSI}$ that treats Classes 1 and 2 as a two-class problem, excluding Class 0. Table \ref{tab:class_confusion_analysis} presents these results, showing how the model's ability to distinguish between Class 1 and Class 2 changed from validation to test data.

\begin{table}[htbp]
    \centering
    \caption{Comparison of classification performance for Class 1 and Class 2 between the validation and test datasets}
    \label{tab:class_confusion_analysis}
    \begin{tabular}{@{}lccccc@{}}
        \toprule
        Region  & \begin{tabular}[c]{@{}c@{}}Validation\\ CSI (A)\end{tabular} & \begin{tabular}[c]{@{}c@{}}Test\\ CSI (B)\end{tabular} & \begin{tabular}[c]{@{}c@{}}Performance\\ Gap ($\Delta$)\tnote{a}\end{tabular} & \begin{tabular}[c]{@{}c@{}}Change\\ (\%)\tnote{b}\end{tabular} \\ \midrule
        Seoul   & 0.6193                                                       & 0.3570                                                 & -0.2623                                                                       & -42.35                                                         \\
        Incheon & 0.6279                                                       & 0.5913                                                 & -0.0366                                                                       & -5.83                                                          \\
        Gwangju & 0.5429                                                       & 0.5070                                                 & -0.0359                                                                       & -6.61                                                          \\
        Daejeon & 0.5361                                                       & 0.4036                                                 & -0.1325                                                                       & -24.71                                                         \\
        Daegu   & 0.4708                                                       & 0.2742                                                 & -0.1967                                                                       & -41.77                                                         \\
        Busan   & 0.5176                                                       & 0.5055                                                 & -0.0121                                                                       & -2.34                                                          \\ \bottomrule
    \end{tabular}
    \begin{tablenotes}
        \item[a] $\Delta = B - A$
        \item[b] $\mathit{Change}(\%) = \dfrac{\Delta}{A} \times 100$
    \end{tablenotes}
\end{table}

We hypothesized that these findings arose from shifts in the distribution of key variables for each class between the train and test datasets, prompting us to conduct further analysis. The variables selected for analysis were those identified through feature importance analysis as having the greatest impact on the predictive model for each region.
Feature importance was calculated using the SHAP value \cite{rozemberczki2022}. For each feature, SHAP values from both Class 0 and Class 1 were summed to derive an overall importance measure. Table \ref{tab:shap_feature_importance} presents the top five variables for each region that exert the greatest influence on the final model and their respective contribution levels.
\begin{table}[htbp]
    \centering
    \footnotesize
    \caption{Top 5 Features by SHAP Importance Share across Regions}
    \label{tab:shap_feature_importance}
    \begin{tabular}{@{}lccccc@{}}
        \toprule
        Region  & \begin{tabular}{c}Top 1\\(IS)\end{tabular}                    & \begin{tabular}{c}Top 2\\(IS)\end{tabular}               & \begin{tabular}{c}Top 3\\(IS)\end{tabular}               & \begin{tabular}{c}Top 4\\(IS)\end{tabular}              & \begin{tabular}{c}Top 5\\(IS)\end{tabular}              \\ \midrule
        Seoul   & \begin{tabular}{c}RH\\(25.33\%)\end{tabular} & \begin{tabular}{c}PM2.5\\(16.65\%)\end{tabular}          & \begin{tabular}{c}dewpoint\_C\\(6.73\%)\end{tabular}     & \begin{tabular}{c}vap\_pressure\\(5.53\%)\end{tabular}  & \begin{tabular}{c}temp\_C\\(4.29\%)\end{tabular}        \\
        Busan   & \begin{tabular}{c}RH\\(18.92\%)\end{tabular} & \begin{tabular}{c}groundtemp\\(8.79\%)\end{tabular}      & \begin{tabular}{c}PM2.5\\(8.19\%)\end{tabular}           & \begin{tabular}{c}temp\_C\\(7.05\%)\end{tabular}        & \begin{tabular}{c}low\_cloudbase\\(5.27\%)\end{tabular} \\
        Incheon & \begin{tabular}{c}RH\\(32.53\%)\end{tabular} & \begin{tabular}{c}cloudcover\\(12.81\%)\end{tabular}     & \begin{tabular}{c}low\_cloudbase\\(9.53\%)\end{tabular}  & \begin{tabular}{c}lm\_cloudcover\\(6.82\%)\end{tabular} & \begin{tabular}{c}PM2.5\\(6.46\%)\end{tabular}          \\
        Daegu   & \begin{tabular}{c}RH\\(23.52\%)\end{tabular} & \begin{tabular}{c}lm\_cloudcover\\(13.18\%)\end{tabular} & \begin{tabular}{c}O3\\(9.32\%)\end{tabular}              & \begin{tabular}{c}PM2.5\\(7.58\%)\end{tabular}          & \begin{tabular}{c}low\_cloudbase\\(7.08\%)\end{tabular} \\
        Daejeon & \begin{tabular}{c}RH\\(20.97\%)\end{tabular} & \begin{tabular}{c}cloudcover\\(13.22\%)\end{tabular}     & \begin{tabular}{c}low\_cloudbase\\(10.26\%)\end{tabular} & \begin{tabular}{c}PM10\\(6.84\%)\end{tabular}           & \begin{tabular}{c}lm\_cloudcover\\(6.80\%)\end{tabular} \\
        Gwangju & \begin{tabular}{c}RH\\(21.81\%)\end{tabular} & \begin{tabular}{c}low\_cloudbase\\(10.10\%)\end{tabular} & \begin{tabular}{c}lm\_cloudcover\\(9.89\%)\end{tabular}  & \begin{tabular}{c}PM2.5\\(7.61\%)\end{tabular}          & \begin{tabular}{c}cloudcover\\(7.28\%)\end{tabular}     \\ \bottomrule
    \end{tabular}
    \begin{tablenotes}
        \item IS: SHAP Importance Share
        \item *Note: For comprehensive SHAP plots covering all regions and features, please refer to the \texttt{shap\_plots} directory in the Supplementary Material repository.
    \end{tablenotes}
\end{table}
We now proceed with an analysis to examine the distribution changes of the top-ranking variables by region. The core issue is that the classification performance for Class 1 and Class 2 in specific regions has sharply changed on the test data compared to the train data.

We hypothesized that for these regions, the distributional distance between training Class 1 and test Class 2 is significantly smaller than the distance between training Class 1 and training Class 2. In other words, the test Class 2 distribution has shifted closer to the training Class 1 distribution. Consequently, the decision boundary learned by the model becomes skewed when applied to the test data.

To measure the distance between distributions, we employ the Wasserstein distance as defined in Section \ref{sec:distributional_distance_metric} (equation \eqref{equation9}). To incorporate feature importance into the distributional distance measurement, we compute a Wasserstein distance focusing on the single most influential feature identified in Table \ref{tab:shap_feature_importance}, which is consistently Relative Humidity (RH) across all regions. We compute the 1-dimensional Wasserstein distance for this top feature individually as shown in equation \eqref{equation10}.

\begin{equation}
    \text{WD}(P_{\text{RH}}^{(A)}, P_{\text{RH}}^{(B)}) = W_1(P_{\text{RH}}^{(A)}, P_{\text{RH}}^{(B)})
    \label{equation10}
\end{equation}

where $W_1(P_{\text{RH}}^{(A)}, P_{\text{RH}}^{(B)})$ denotes the 1-dimensional Wasserstein distance between the marginal distributions of the Relative Humidity feature in dataset A and dataset B, respectively. This process allows us to quantify distributional shifts while prioritizing the feature that contributes most to model predictions.
The following are the comparison targets for each region using Wasserstein distance (equations \eqref{equation11} and \eqref{equation12}).

\begin{align}
    D_{\mathrm{base}}
     & = \text{WD}\bigl(P^{\mathrm{train}}_{1},\, P^{\mathrm{train}}_{2}\bigr) \label{equation11} \\
    D_{\mathrm{shift}}
     & = \text{WD}\bigl(P^{\mathrm{train}}_{1},\, P^{\mathrm{test}}_{2}\bigr) \label{equation12}
\end{align}

where $P^{\mathrm{train}}_{c}$ and $P^{\mathrm{test}}_{c}$ denote the class-conditional distributions of the features given $\mathrm{Target}=\mathrm{Class\ }c$ in the training and test sets, respectively.

Table \ref{tab:wasserstein_rh_comparison} presents the results of calculating and comparing the two distance measures for the most influential feature in each region. Specifically, it shows the Wasserstein distance computed for the Relative Humidity variable, which consistently ranked as the top feature by SHAP importance across all six regions. For detailed per-feature Wasserstein distance analysis for other high-importance variables, see Appendix \ref{app:per_feature_wd}.

\begin{table}[htbp]
    \centering
    \caption{Inter-class Wasserstein Distance Analysis for RH ($D_{\mathrm{base}}$ vs $D_{\mathrm{shift}}$)}
    \label{tab:wasserstein_rh_comparison}
    \begin{tabular}{lccc}
        \toprule
        Region  & $\mathrm{D}_{\mathrm{base}}$ & $\mathrm{D}_{\mathrm{shift}}$ & Change (\%)\tnote{a} \\ \midrule
        Seoul   & 1.2686                       & 0.9010                        & -28.97               \\
        Busan   & 1.5029                       & 1.3798                        & -8.19                \\
        Incheon & 1.2590                       & 1.4222                        & +12.96               \\
        Daegu   & 1.2869                       & 1.0505                        & -18.37               \\
        Daejeon & 1.0699                       & 0.9167                        & -14.31               \\
        Gwangju & 1.0410                       & 0.9123                        & -12.36               \\ \bottomrule
    \end{tabular}
    \begin{tablenotes}
        \item[a] $\mathit{Change}(\%) = \dfrac{D_{\mathrm{shift}} - D_{\mathrm{base}}}{D_{\mathrm{base}}} \times 100$
    \end{tablenotes}
\end{table}

For regions exhibiting severe performance degradation (Seoul, Daegu, Daejeon), we observed that $D_{\mathrm{shift}}$ was significantly closer than $D_{\mathrm{base}}$. That is, the distance between the Class 1 distribution in the validation data and the Class 2 distribution in the test data decreased significantly compared to the original distance between the Class 1-2 distributions within the validation data. This indicates that the distribution of Class 2 in the test data shifted substantially towards the Class 1 distribution, suggesting a sharp change in the discriminative power of the learned decision boundary.

In regions with negligible performance degradation (Busan, Incheon, Gwangju), we observed that $D_{\mathrm{shift}}$ was either close to $D_{\mathrm{base}}$ or even considerably farther. That is, the relative distance relationship between Class distributions in the test data sufficiently maintained the structure observed in the validation data, and thus the learned decision boundary functioned effectively on the test data as well.

These results strongly support the hypothesis that distribution shift invalidated the decision boundary between Class 1 and Class 2.

Beyond the distribution shift analysis, we also examined the model's performance on the most severe low-visibility class. The confusion matrix in Appendix \ref{app:confusion_matrices} reveals that the true positive rate for Class 0 is exceptionally low in certain regions. However, given the extremely small number of actual Class 0 data points in those regions, it cannot be definitively stated that the model's classification performance for Class 0 is low. Indeed, there is an established perspective that when data counts are extremely low, the reliability of metrics is also not high \cite{briscoe2025}. Furthermore, for our model's learning metric, multiple CSI, the data size for Class 0 is overwhelmingly smaller than that for Class 1. Therefore, the possibility that the model was trained prioritizing the generalization performance of Class 1 cannot be ruled out \cite{francazi2023}.

\subsection{Baseline Comparison}\label{sec:baseline_comparison}

Having completed our analysis of the reasons for the change in the test performance of our final model in certain regions, we shall finally determine whether the predictive performance of the final model has improved by comparing it with that of the existing traditional statistical model. Table \ref{tab:baseline_comparison} compares the validation performance and test performance of the final selected model in this study with those of the traditional statistical model, the logistic regression model. This table thus provides a comparison of the test prediction performance and the generalization performance on the validation data and test data.

\begin{table}[htbp]
    \centering
    \caption{Comparison of Test Performance between Logistic Regression and Ensemble Model by Region}
    \label{tab:baseline_comparison}
    \begin{tabular}{lccccccccc}
        \toprule
        \multirow{2}{*}{Region} & \multicolumn{3}{c}{Logistic Regression} & \multicolumn{3}{c}{Ensemble Model} & \multicolumn{3}{c}{Change(\%)}                                                                         \\ \cmidrule(lr){2-10}
                                & CSI                                     & MCC                                & ACC                            & CSI    & MCC    & ACC    & CSI\tnote{a} & MCC\tnote{b} & ACC\tnote{c} \\ \midrule
        Seoul                   & 0.3820                                  & 0.5463                             & 0.9342                         & 0.3521 & 0.5415 & 0.9092 & -8.4913      & -0.8984      & -2.7495      \\
        Incheon                 & 0.5178                                  & 0.6674                             & 0.9182                         & 0.5615 & 0.6947 & 0.9227 & 7.7836       & 3.9343       & 0.4949       \\
        Gwangju                 & 0.4494                                  & 0.6086                             & 0.9497                         & 0.5052 & 0.6487 & 0.9459 & 11.0410      & 6.1775       & -0.3983      \\
        Daejeon                 & 0.2803                                  & 0.4161                             & 0.9153                         & 0.4078 & 0.5687 & 0.9164 & 31.2572      & 26.8298      & 0.1246       \\
        Daegu                   & 0.2655                                  & 0.4121                             & 0.9811                         & 0.2739 & 0.4527 & 0.9740 & 3.0665       & 8.9601       & -0.7267      \\
        Busan                   & 0.3822                                  & 0.5545                             & 0.9692                         & 0.4712 & 0.6541 & 0.9707 & 18.8972      & 15.2267      & 0.1529       \\ \bottomrule
    \end{tabular}
    \begin{tablenotes}
        \setlength{\itemsep}{1pt}
        \item[a] $CSI(\mathrm{\%}) = \dfrac{\text{Ensemble}_{CSI} -\text{Logistic}_{CSI}}{\text{Ensemble}_{CSI}} \times 100$
        \item[b] $MCC(\mathrm{\%}) = \dfrac{\text{Ensemble}_{MCC} -\text{Logistic}_{MCC}}{\text{Ensemble}_{MCC}} \times 100$
        \item[c] $ACC(\mathrm{\%}) = \dfrac{\text{Ensemble}_{ACC} -\text{Logistic}_{ACC}}{\text{Ensemble}_{ACC}} \times 100$
    \end{tablenotes}
\end{table}

Table \ref{tab:baseline_comparison} compares the test performance between the logistic regression and our ensemble model. Our ensemble model outperformed the logistic regression baseline in five of the six regions, with Seoul being the only exception where the logistic regression showed slightly better performance.

\section{Conclusion}\label{sec:conclusion}

This study presented a machine learning methodology to tackle the dual challenges of data imbalance and distribution shift in visibility nowcasting for six major cities in South Korea. We employed SMOTENC and CTGAN to alleviate severe class imbalance and evaluated five machine learning and deep learning models along with a two-stage ensemble technique.

Our analysis yielded several key findings. First, through UMAP visualization, Wasserstein distance, and KDE analysis, we verified that the synthetic data generated by each augmentation method generally preserved the distributional characteristics of the original data, though with varying degrees of fidelity. Second, the optimal augmentation strategy differed by model architecture: tree-based models such as XGBoost achieved the greatest performance improvement with SMOTENC, while deep learning models such as ResNet-like performed better with CTGAN augmentation. This finding suggests that distributional fidelity does not necessarily translate to better classification performance. Third, our ensemble model outperformed the logistic regression baseline in five of the six regions, demonstrating the potential of the proposed approach for visibility nowcasting. However, the ensemble technique did not uniformly improve performance across all regions, indicating that its effectiveness depends on the specific characteristics of each dataset.

This study also identified important limitations. A significant performance degradation was observed in certain regions (Seoul, Daegu, Daejeon) when evaluated on test data, which we attributed to distribution shift between training and test periods. Additionally, the extremely low prevalence of Class 0 (severe low-visibility events) made it difficult to reliably assess the model's performance for this critical class.

Future research should address these limitations through several directions: (1) securing multi-year datasets to improve model generalization across temporal variations, (2) collecting higher-resolution data (e.g., 10-minute intervals) to better capture the dynamic characteristics of visibility-related variables, and (3) implementing high-resolution data collection strategies to reliably obtain low-prevalence event data.

In summary, this study emphasizes that for securing the long-term reliability of visibility prediction models, continuous monitoring of data distribution changes and adaptive model updating strategies are essential. These findings are expected to contribute to the development of more accurate and robust visibility nowcasting systems, thereby enhancing safety in air, sea, and road transportation and supporting related environmental risk management \cite{smith2023}.

\backmatter

\section*{Supplementary Information}
The online version contains supplementary material available at \url{https://doi.org/10.6084/m9.figshare.31252024}.

\section{Declarations}

\subsection{Funding}
This research received no external funding.

\subsection{Competing interests}
The authors have no relevant financial or nonfinancial interests to disclose.

\subsection{Ethics approval and consent to participate}
Not applicable.

\subsection{Consent for publication}
Not applicable.

\subsection{Data availability}
The datasets analyzed for this study can be found in the public repositories cited. Specifically, meteorological data from the Automated Synoptic Observing System (ASOS) are publicly available from the Korea Meteorological Administration (KMA) at \url{https://data.kma.go.kr/data/grnd/selectAsosRltmList.do} (accessed on 26 August 2025). Air quality data are available through the AirKorea website, operated by the Korea Environment Corporation, at \url{https://www.airkorea.or.kr/web/last\_amb_hour\_data?pMENU\_NO=123} (accessed on 26 August 2025). The supplementary materials, including hyperparameter optimization logs, SHAP plots, and detailed model configurations, are available at Figshare: \url{https://doi.org/10.6084/m9.figshare.31252024}.

\subsection{Code availability}
The code generated during the current study is available in the GitHub repository, \url{https://github.com/singbong/Visibility_Nowcasting}.

\subsection{Author contributions}
Conceptualization, H.S.S.; methodology, H.S.S.; supervision, H.S.S.; writing—review and editing, H.S.S. Data curation, B.G.S. and C.S.L.; formal analysis, B.G.S. and C.S.L.; software, B.G.S. and C.S.L.; writing—original draft preparation, B.G.S. and C.S.L. B.G.S. and C.S.L. contributed equally to this work. B.G.S. was solely responsible for communication with the corresponding author. All authors have read and agreed to the published version of the manuscript.
\clearpage
\begin{appendices}
    \section{Best \texorpdfstring{$\mathrm{CSI}$}{CSI} Score by Augmentation Model for each Region}\label{app:best_csi_by_model}
    The following figures present detailed CSI performance comparisons across all augmentation methods for each prediction model and region, complementing the aggregate results shown in Figure \ref{fig:csi_improvement} in the main text.
    \begin{figure}[H]
    \centering
        \includegraphics[width=0.85\textwidth]{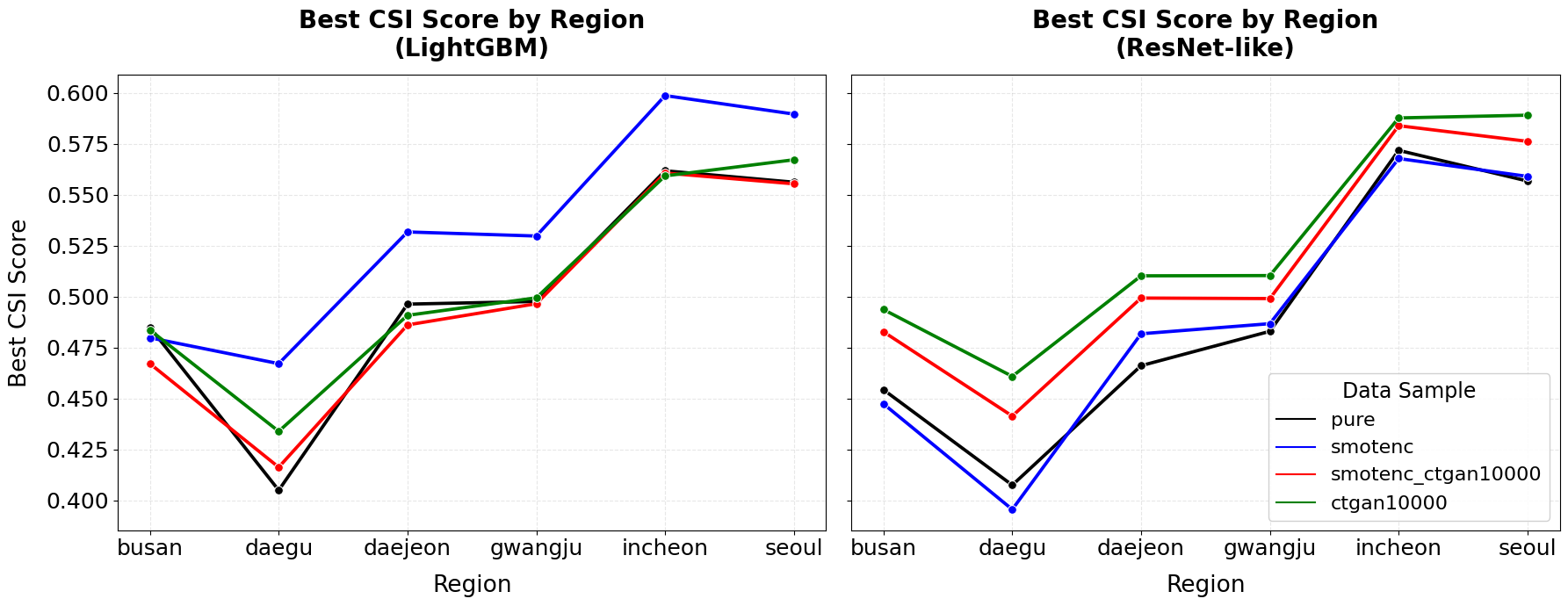}
        \caption{\small{Best CSI performance by augmentation method: LightGBM and ResNet-like models}}
        \label{fig:app_lightgbm_resnet}
        \vspace{0.2cm}
        \includegraphics[width=0.85\textwidth]{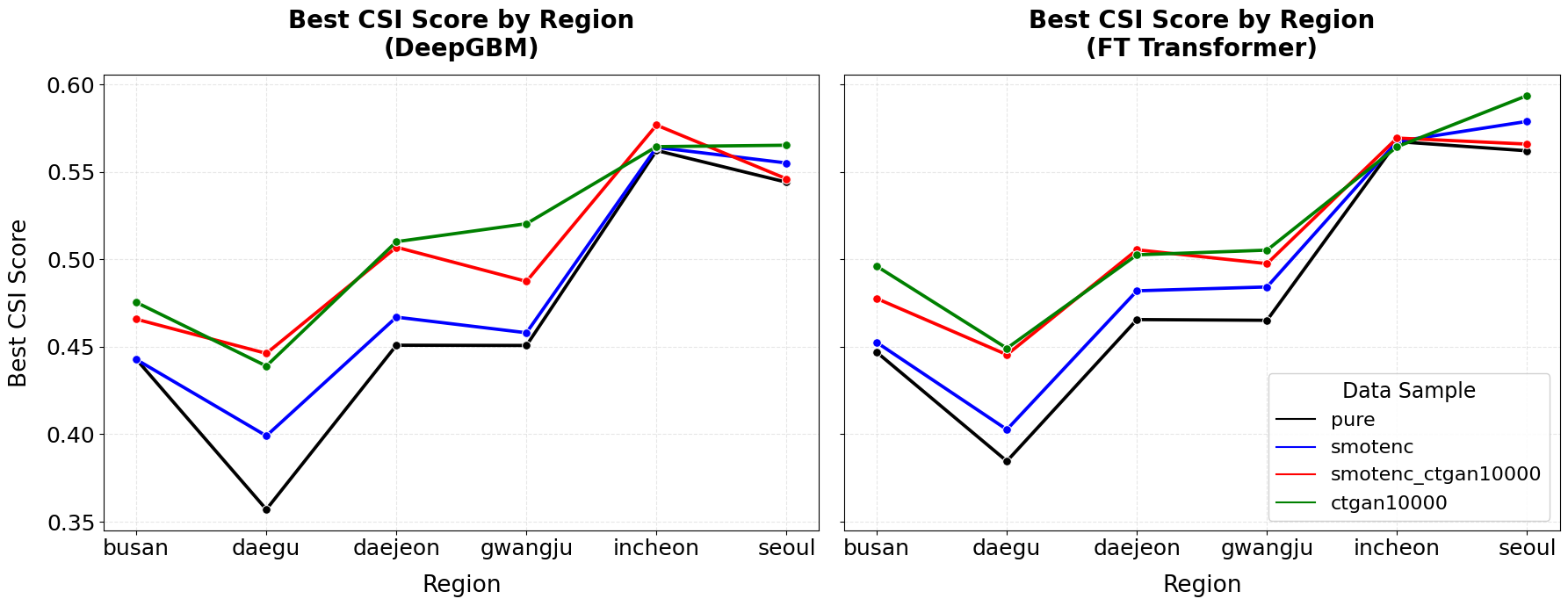}
        \caption{\small{Best CSI performance by augmentation method: DeepGBM and FT-Transformer models}}
        \label{fig:app_deepgbm_ftt}
        
        \vspace{0.2cm}
        
        \includegraphics[width=0.45\textwidth]{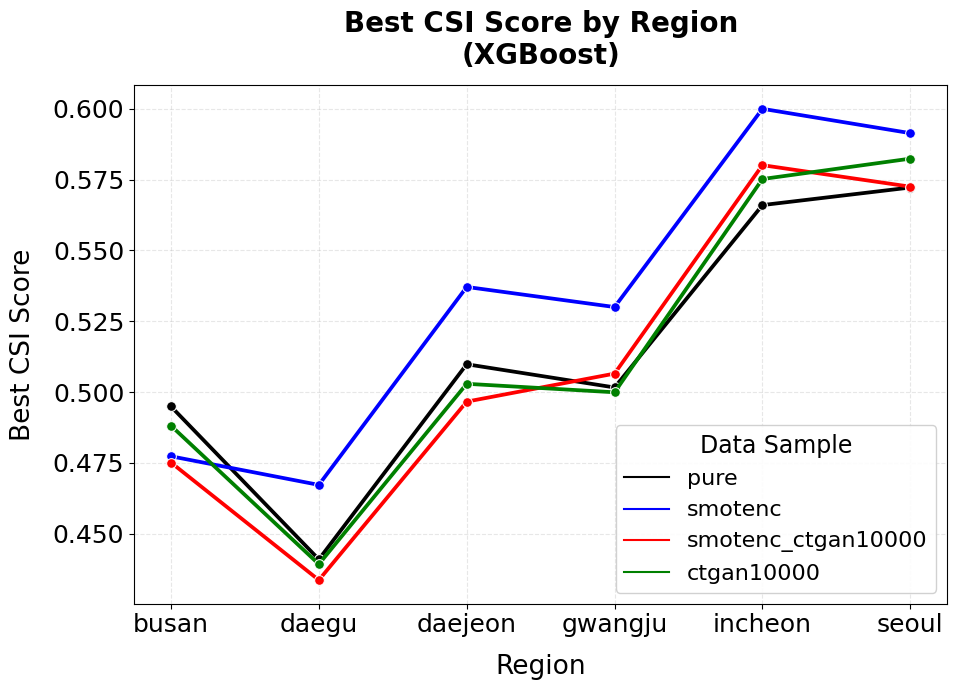}
        \caption{\small{Best CSI performance by augmentation method: XGBoost model}}
        \label{fig:app_xgboost}
    \end{figure}

    \clearpage
    
    \section{Confusion Matrices for All Regions}\label{app:confusion_matrices}

    The following tables present the confusion matrices from the final ensemble model predictions on the test dataset for each region.
    
    \begin{table}[htbp]
        \centering
        \begin{minipage}[t]{0.48\textwidth}
            \centering
            \resizebox{\linewidth}{!}{%
            \footnotesize
            \setlength{\tabcolsep}{5pt}
            \renewcommand{\arraystretch}{1.1}
            \begin{tabular}{@{\hspace{6pt}}lccc@{\hspace{6pt}}}
                \toprule
                & Pred 0 & Pred 1 & Pred 2 \\
                \midrule
                True 0 & \cellcolor{blue!25}0  & 13 & 0   \\
                True 1 & 2 & \cellcolor{blue!25}432 & 47  \\
                True 2 & 2 & 731 & \cellcolor{blue!25}7533 \\
                \bottomrule
            \end{tabular}%
            }
            \captionof{table}{\small{Confusion matrix for Seoul test dataset}}
            \label{tab:confusion_seoul_test}
        \end{minipage}
        \hfill
        \begin{minipage}[t]{0.48\textwidth}
            \centering
            \resizebox{\linewidth}{!}{%
            \footnotesize
            \setlength{\tabcolsep}{5pt}
            \renewcommand{\arraystretch}{1.1}
            \begin{tabular}{@{\hspace{6pt}}lccc@{\hspace{6pt}}}
                \toprule
                & Pred 0 & Pred 1 & Pred 2 \\
                \midrule
                True 0 & \cellcolor{blue!25}0   & 24 & 0   \\
                True 1 & 8 & \cellcolor{blue!25}229 & 44  \\
                True 2 & 1 & 180 & \cellcolor{blue!25}8274 \\
                \bottomrule
            \end{tabular}%
            }
            \captionof{table}{\small{Confusion matrix for Busan test dataset}}
            \label{tab:confusion_busan_test}
        \end{minipage}
    \end{table}
    \begin{table}[htbp]
        \centering
        \begin{minipage}[t]{0.48\textwidth}
            \centering
            \resizebox{\linewidth}{!}{%
            \footnotesize
            \setlength{\tabcolsep}{5pt}
            \renewcommand{\arraystretch}{1.1}
            \begin{tabular}{@{\hspace{6pt}}lccc@{\hspace{6pt}}}
                \toprule
                & Pred 0 & Pred 1 & Pred 2 \\
                \midrule
                True 0 & \cellcolor{blue!25}64  & 86 & 32   \\
                True 1 & 4 & \cellcolor{blue!25}803 & 398  \\
                True 2 & 0 & 157 & \cellcolor{blue!25}7216 \\
                \bottomrule
            \end{tabular}%
            }
            \captionof{table}{\small{Confusion matrix for Incheon test dataset}}
            \label{tab:confusion_incheon_test}
        \end{minipage}
        \hfill
        \begin{minipage}[t]{0.48\textwidth}
            \centering
            \resizebox{\linewidth}{!}{%
            \footnotesize
            \setlength{\tabcolsep}{5pt}
            \renewcommand{\arraystretch}{1.1}
            \begin{tabular}{@{\hspace{6pt}}lccc@{\hspace{6pt}}}
                \toprule
                & Pred 0 & Pred 1 & Pred 2 \\
                \midrule
                True 0 & \cellcolor{blue!25}1   & 0 & 0   \\
                True 1 & 2 & \cellcolor{blue!25}85 & 41  \\
                True 2 & 1 & 184 & \cellcolor{blue!25}8446 \\
                \bottomrule
            \end{tabular}%
            }
            \captionof{table}{\small{Confusion matrix for Daegu test dataset}}
            \label{tab:confusion_daegu_test}
        \end{minipage}
    \end{table}
    \begin{table}[htbp]
        \centering
        \begin{minipage}[t]{0.48\textwidth}
            \centering
            \resizebox{\linewidth}{!}{%
            \footnotesize
            \setlength{\tabcolsep}{5pt}
            \renewcommand{\arraystretch}{1.1}
            \begin{tabular}{@{\hspace{6pt}}lccc@{\hspace{6pt}}}
                \toprule
                & Pred 0 & Pred 1 & Pred 2 \\
                \midrule
                True 0 & \cellcolor{blue!25}35  & 16 & 2   \\
                True 1 & 21 & \cellcolor{blue!25}469 & 128  \\
                True 2 & 0 & 565 & \cellcolor{blue!25}7524 \\
                \bottomrule
            \end{tabular}%
            }
            \captionof{table}{\small{Confusion matrix for Daejeon test dataset}}
            \label{tab:confusion_daejeon_test}
        \end{minipage}
        \hfill
        \begin{minipage}[t]{0.48\textwidth}
            \centering
            \resizebox{\linewidth}{!}{%
            \footnotesize
            \setlength{\tabcolsep}{5pt}
            \renewcommand{\arraystretch}{1.1}
            \begin{tabular}{@{\hspace{6pt}}lccc@{\hspace{6pt}}}
                \toprule
                & Pred 0 & Pred 1 & Pred 2 \\
                \midrule
                True 0 & \cellcolor{blue!25}15  & 12 & 3   \\
                True 1 & 3 & \cellcolor{blue!25}469 & 171  \\
                True 2 & 0 & 285 & \cellcolor{blue!25}7802 \\
                \bottomrule
            \end{tabular}%
            }
            \captionof{table}{\small{Confusion matrix for Gwangju test dataset}}
            \label{tab:confusion_gwangju_test}
        \end{minipage}
    \end{table}
    \clearpage

    \section{Per-Feature Wasserstein Distance Analysis for All Regions}\label{app:per_feature_wd}
    \setcounter{table}{0}

    The following tables present per-feature Wasserstein Distance analysis for each region, where IS denotes SHAP Importance Share and Change (\%) denotes the percentage change from $D_{\mathrm{base}}$ to $D_{\mathrm{shift}}$. For the detailed SHAP plots corresponding to these features, please refer to the \texttt{shap\_plots} directory in the Supplementary Material repository.

        \begin{table}[htbp]
        \centering
        \begin{minipage}[t]{0.48\textwidth}
            \centering
            \resizebox{\linewidth}{!}{%
            \small
            \setlength{\tabcolsep}{4pt}
            \renewcommand{\arraystretch}{1.05}
            \begin{tabular}{@{\hspace{4pt}}lcccc@{\hspace{4pt}}}
                \toprule
                Feature       & IS & $\mathrm{D}_{\mathrm{base}}$ & $\mathrm{D}_{\mathrm{shift}}$ & Change (\%) \\ \midrule
                RH            & 0.253       & 1.2686                       & 0.9010                        & -28.97               \\
                PM2.5          & 0.167       & 1.4134                       & 1.4598                        & 3.29                 \\
                dewpoint\_C   & 0.067       & 0.3865                       & 0.2006                        & -48.09               \\
                vap\_pressure & 0.055       & 0.2890                       & 0.1376                        & -52.39               \\
                temp\_C       & 0.043       & 0.2543                       & 0.2873                        & 12.96                \\ \bottomrule
            \end{tabular}%
            }
            \captionof{table}{\small{Wasserstein distance between Class 1 and Class 2 distributions in Seoul}}
        \end{minipage}
        \hfill
        \begin{minipage}[t]{0.48\textwidth}
            \centering
            \resizebox{\linewidth}{!}{%
            \small
            \setlength{\tabcolsep}{4pt}
            \renewcommand{\arraystretch}{1.05}
            \begin{tabular}{@{\hspace{4pt}}lcccc@{\hspace{4pt}}}
                \toprule
                Feature        & IS & $\mathrm{D}_{\mathrm{base}}$ & $\mathrm{D}_{\mathrm{shift}}$ & Change (\%) \\ \midrule
                RH             & 0.189       & 1.5029                       & 1.3798                        & -8.19                \\
                groundtemp     & 0.088       & 0.4254                       & 0.3890                        & -8.56                \\
                PM25           & 0.082       & 0.3188                       & 0.6326                        & 98.41                \\
                temp\_C        & 0.071       & 0.4209                       & 0.3799                        & -9.73                \\
                low\_cloudbase & 0.053       & 1.0656                       & 0.9453                        & -11.29               \\ \bottomrule
            \end{tabular}%
            }
            \captionof{table}{\small{Wasserstein distance between Class 1 and Class 2 distributions in Busan}}
        \end{minipage}
    \end{table}
    \begin{table}[htbp]
        \centering
        \begin{minipage}[t]{0.48\textwidth}
            \centering
            \resizebox{\linewidth}{!}{%
            \small
            \setlength{\tabcolsep}{4pt}
            \renewcommand{\arraystretch}{1.05}
            \begin{tabular}{@{\hspace{4pt}}lcccc@{\hspace{4pt}}}
                \toprule
                Feature        & IS & $\mathrm{D}_{\mathrm{base}}$ & $\mathrm{D}_{\mathrm{shift}}$ & Change (\%) \\ \midrule
                RH             & 0.325       & 1.2590                       & 1.4222                        & 12.96                \\
                cloudcover     & 0.128       & 0.4649                       & 0.4104                        & -11.72               \\
                low\_cloudbase & 0.095       & 0.4986                       & 0.4556                        & -8.61                \\
                lm\_cloudcover & 0.068       & 0.5873                       & 0.5895                        & 0.37                 \\
                PM25           & 0.065       & 1.2347                       & 1.3226                        & 7.12                 \\ \bottomrule
            \end{tabular}%
            }
            \captionof{table}{\small{Wasserstein distance between Class 1 and Class 2 distributions in Incheon}}
        \end{minipage}
        \hfill
        \begin{minipage}[t]{0.48\textwidth}
            \centering
            \resizebox{\linewidth}{!}{%
            \small
            \setlength{\tabcolsep}{4pt}
            \renewcommand{\arraystretch}{1.05}
            \begin{tabular}{@{\hspace{4pt}}lcccc@{\hspace{4pt}}}
                \toprule
                Feature        & IS & $\mathrm{D}_{\mathrm{base}}$ & $\mathrm{D}_{\mathrm{shift}}$ & Change (\%) \\ \midrule
                RH             & 0.235       & 1.2869                       & 1.0505                        & -18.37               \\
                lm\_cloudcover & 0.132       & 0.6347                       & 0.6107                        & -3.78                \\
                O3             & 0.093       & 0.8072                       & 0.8323                        & 3.11                 \\
                PM25           & 0.076       & 1.5369                       & 1.7213                        & 12.00                \\
                low\_cloudbase & 0.071       & 0.4924                       & 0.4458                        & -9.46                \\ \bottomrule
            \end{tabular}%
            }
            \captionof{table}{\small{Wasserstein distance between Class 1 and Class 2 distributions in Daegu}}
        \end{minipage}
    \end{table}

    \begin{table}[htbp]
        \centering
        \begin{minipage}[t]{0.48\textwidth}
            \centering
            \resizebox{\linewidth}{!}{%
            \small
            \setlength{\tabcolsep}{4pt}
            \renewcommand{\arraystretch}{1.05}
            \begin{tabular}{@{\hspace{4pt}}lcccc@{\hspace{4pt}}}
                \toprule
                Feature        & IS & $\mathrm{D}_{\mathrm{base}}$ & $\mathrm{D}_{\mathrm{shift}}$ & Change (\%) \\ \midrule
                RH             & 0.210       & 1.0699                       & 0.9167                        & -14.31               \\
                cloudcover     & 0.132       & 0.4627                       & 0.3464                        & -25.14               \\
                low\_cloudbase & 0.103       & 0.3744                       & 0.3624                        & -3.18                \\
                PM10           & 0.068       & 0.8017                       & 0.9233                        & 15.18                \\
                lm\_cloudcover & 0.068       & 0.4922                       & 0.4490                        & -8.78                \\ \bottomrule
            \end{tabular}%
            }
            \captionof{table}{\small{Wasserstein distance between Class 1 and Class 2 distributions in Daejeon}}
        \end{minipage}
        \hfill
        \begin{minipage}[t]{0.48\textwidth}
            \centering
            \resizebox{\linewidth}{!}{%
            \small
            \setlength{\tabcolsep}{4pt}
            \renewcommand{\arraystretch}{1.05}
            \begin{tabular}{@{\hspace{4pt}}lcccc@{\hspace{4pt}}}
                \toprule
                Feature        & IS & $\mathrm{D}_{\mathrm{base}}$ & $\mathrm{D}_{\mathrm{shift}}$ & Change (\%) \\ \midrule
                RH             & 0.218       & 1.0410                       & 0.9123                        & -12.36               \\
                low\_cloudbase & 0.101       & 0.4350                       & 0.4611                        & 6.02                 \\
                lm\_cloudcover & 0.099       & 0.4314                       & 0.4652                        & 7.83                 \\
                PM25           & 0.076       & 1.2919                       & 1.4873                        & 15.13                \\
                cloudcover     & 0.073       & 0.4333                       & 0.3305                        & -23.72               \\ \bottomrule
            \end{tabular}%
            }
            \captionof{table}{\small{Wasserstein distance between Class 1 and Class 2 distributions in Gwangju}}
        \end{minipage}
    \end{table}

    \begin{table}[htbp]
        \centering
        \begin{minipage}[t]{0.48\textwidth}
            \centering
            \resizebox{\linewidth}{!}{%
            \small
            \setlength{\tabcolsep}{4pt}
            \renewcommand{\arraystretch}{1.05}
            \begin{tabular}{@{\hspace{4pt}}lcccc@{\hspace{4pt}}}
                \toprule
                Feature  & IS & $\mathrm{D}_{\mathrm{base}}$ & $\mathrm{D}_{\mathrm{shift}}$ & Change (\%) \\ \midrule
                PM2.5     & 0.144       & 0.4388                       & 0.5174                        & 17.93                \\
                PM10     & 0.084       & 0.4446                       & 1.2402                        & 178.93               \\
                solarRad & 0.083       & 0.3994                       & 0.2734                        & -31.56               \\
                temp\_C  & 0.078       & 0.6918                       & 0.6674                        & -3.52                \\
                hm       & 0.073       & 0.9138                       & 0.3542                        & -61.24               \\ \bottomrule
            \end{tabular}%
            }
            \captionof{table}{\small{Wasserstein distance between Class 0 and Class 1 distributions in Seoul}}
        \end{minipage}
        \hfill
        \begin{minipage}[t]{0.48\textwidth}
            \centering
            \resizebox{\linewidth}{!}{%
            \small
            \setlength{\tabcolsep}{4pt}
            \renewcommand{\arraystretch}{1.05}
            \begin{tabular}{@{\hspace{4pt}}lcccc@{\hspace{4pt}}}
                \toprule
                Feature       & IS & $\mathrm{D}_{\mathrm{base}}$ & $\mathrm{D}_{\mathrm{shift}}$ & Change (\%) \\ \midrule
                RH            & 0.227       & 0.4986                       & 0.3144                        & -36.93               \\
                temp\_C       & 0.084       & 0.3816                       & 0.3500                        & -8.28                \\
                PM2.5          & 0.068       & 0.3790                       & 0.4033                        & 6.42                 \\
                dewpoint\_C   & 0.064       & 0.4053                       & 0.3893                        & -3.95                \\
                vap\_pressure & 0.061       & 0.3799                       & 0.4029                        & 6.08                 \\ \bottomrule
            \end{tabular}%
            }
            \captionof{table}{\small{Wasserstein distance between Class 0 and Class 1 distributions in Busan}}
        \end{minipage}
    \end{table}

    \begin{table}[htbp]
        \centering
        \begin{minipage}[t]{0.48\textwidth}
            \centering
            \resizebox{\linewidth}{!}{%
            \small
            \setlength{\tabcolsep}{4pt}
            \renewcommand{\arraystretch}{1.05}
            \begin{tabular}{@{\hspace{4pt}}lcccc@{\hspace{4pt}}}
                \toprule
                Feature        & IS & $\mathrm{D}_{\mathrm{base}}$ & $\mathrm{D}_{\mathrm{shift}}$ & Change (\%) \\ \midrule
                RH             & 0.270       & 0.9400                       & 1.3852                        & 47.37                \\
                cloudcover     & 0.159       & 0.8924                       & 0.9215                        & 3.27                 \\
                low\_cloudbase & 0.110       & 0.7946                       & 0.8863                        & 11.55                \\
                lm\_cloudcover & 0.101       & 1.1641                       & 1.2814                        & 10.08                \\
                solarRad       & 0.048       & 0.4238                       & 0.4761                        & 12.34                \\ \bottomrule
            \end{tabular}%
            }
            \captionof{table}{\small{Wasserstein distance between Class 0 and Class 1 distributions in Incheon}}
        \end{minipage}
        \hfill
        \begin{minipage}[t]{0.48\textwidth}
            \centering
            \resizebox{\linewidth}{!}{%
            \small
            \setlength{\tabcolsep}{4pt}
            \renewcommand{\arraystretch}{1.05}
            \begin{tabular}{@{\hspace{4pt}}lcccc@{\hspace{4pt}}}
                \toprule
                Feature        & IS & $\mathrm{D}_{\mathrm{base}}$ & $\mathrm{D}_{\mathrm{shift}}$ & Change (\%) \\ \midrule
                lm\_cloudcover & 0.239       & 1.2692                       & 0.9963                        & -21.50               \\
                RH             & 0.173       & 0.7080                       & 0.2719                        & -61.60               \\
                O$_3$          & 0.142       & 0.7597                       & 0.6814                        & -10.29               \\
                low\_cloudbase & 0.109       & 0.7863                       & 0.4078                        & -48.14               \\
                hour\_sin      & 0.092       & 0.7369                       & 0.6374                        & -13.51               \\ \bottomrule
            \end{tabular}%
            }
            \captionof{table}{\small{Wasserstein distance between Class 0 and Class 1 distributions in Daegu}}
        \end{minipage}
    \end{table}

    \begin{table}[htbp]
        \centering
        \begin{minipage}[t]{0.48\textwidth}
            \centering
            \resizebox{\linewidth}{!}{%
            \small
            \setlength{\tabcolsep}{4pt}
            \renewcommand{\arraystretch}{1.05}
            \begin{tabular}{@{\hspace{4pt}}lcccc@{\hspace{4pt}}}
                \toprule
                Feature        & IS & $\mathrm{D}_{\mathrm{base}}$ & $\mathrm{D}_{\mathrm{shift}}$ & Change (\%) \\ \midrule
                cloudcover     & 0.188       & 0.8681                       & 0.8688                        & 0.08                 \\
                low\_cloudbase & 0.137       & 0.8057                       & 0.7877                        & -2.24                \\
                RH             & 0.113       & 0.6220                       & 0.2091                        & -66.39               \\
                lm\_cloudcover & 0.099       & 1.0474                       & 0.9909                        & -5.39                \\
                PM10           & 0.062       & 0.2931                       & 0.6929                        & 136.35               \\ \bottomrule
            \end{tabular}%
            }
            \captionof{table}{\small{Wasserstein distance between Class 0 and Class 1 distributions in Daejeon}}
        \end{minipage}
        \hfill
        \begin{minipage}[t]{0.48\textwidth}
            \centering
            \resizebox{\linewidth}{!}{%
            \small
            \setlength{\tabcolsep}{4pt}
            \renewcommand{\arraystretch}{1.05}
            \begin{tabular}{@{\hspace{4pt}}lcccc@{\hspace{4pt}}}
                \toprule
                Feature        & IS & $\mathrm{D}_{\mathrm{base}}$ & $\mathrm{D}_{\mathrm{shift}}$ & Change (\%) \\ \midrule
                lm\_cloudcover & 0.148       & 1.0935                       & 1.0019                        & -8.38                \\
                RH             & 0.131       & 0.5440                       & 0.3275                        & -39.81               \\
                low\_cloudbase & 0.129       & 0.6667                       & 0.6545                        & -1.84                \\
                cloudcover     & 0.081       & 0.7659                       & 0.6409                        & -16.31               \\
                precip\_mm     & 0.064       & 0.8630                       & 0.8917                        & 3.33                 \\ \bottomrule
            \end{tabular}%
            }
            \captionof{table}{\small{Wasserstein distance between Class 0 and Class 1 distributions in Gwangju}}
        \end{minipage}
    \end{table}

\end{appendices}
\clearpage
\bibliography{sn-bibliography}% common bib file
\end{document}